# Law-Invariant Operators for Stochastic Mechanics

*Vissarion Papadopoulos*

Institute of Structural Analysis and Seismic Research,
National Technical University of Athens,
Iroon Polytechniou 9, Zografou Campus, Athens 15780, Greece
Email: vpapado@central.ntua.gr

---

**Abstract**

A central objective in stochastic mechanics is to separate deterministic physical behavior from the probability laws governing uncertain inputs. While classical spectral approaches achieve this separation in linear settings, nonlinear responses introduce interactions beyond conventional second-order variability response functions (VRF). Inspired by the spirit of VRFs, this work establishes that, for a fixed physical system, uncertainty propagation is described by a law-invariant operator mapping input realizations to responses, governed by the underlying physics and the deterministic parameters of the system. Different probability laws affect the input sampling and resulting response statistics, but not the mapping operator. We establish this law invariance under complete, invertible coordinate transformations and develop complementary, Fourier, fixed-reference polynomial-chaos (PCE) and finite element method (FEM) transfer formulations. Fourier captures nonlinear cross-frequency interactions and recovers the classical VRF in the appropriate linear limit, PCE describes coupling among stochastic modes, while FEM expresses the deterministic inverse-stiffness action. Neural approximations of these deterministic operators are investigated through analytical and finite-element benchmarks, examining cross-law generalization, the influence of training distributions and stochastic excitation, and the prediction of response fields and statistics. The results demonstrate the potential for reusing learned physical operators across different stochastic laws without retraining, while highlighting the importance of adequate training coverage and the challenges of accurately estimating low-probability events.

## Highlights

- The deterministic solution operator is separated from the probability law of its stochastic input.
- FEM inverse-stiffness, Fourier and fixed-reference PCE formulations describe the same law-invariant underlying physics.
- Classical VRF variance transfer is recovered in the appropriate linear limit.
- Cross-law transfer is evaluated for physical and Fourier-coordinate finite-element surrogates under unseen modulus laws.
- Stochastic training measures and neural architectures influence operator accuracy and rare-event probability predictions.

## Graphical Abstract

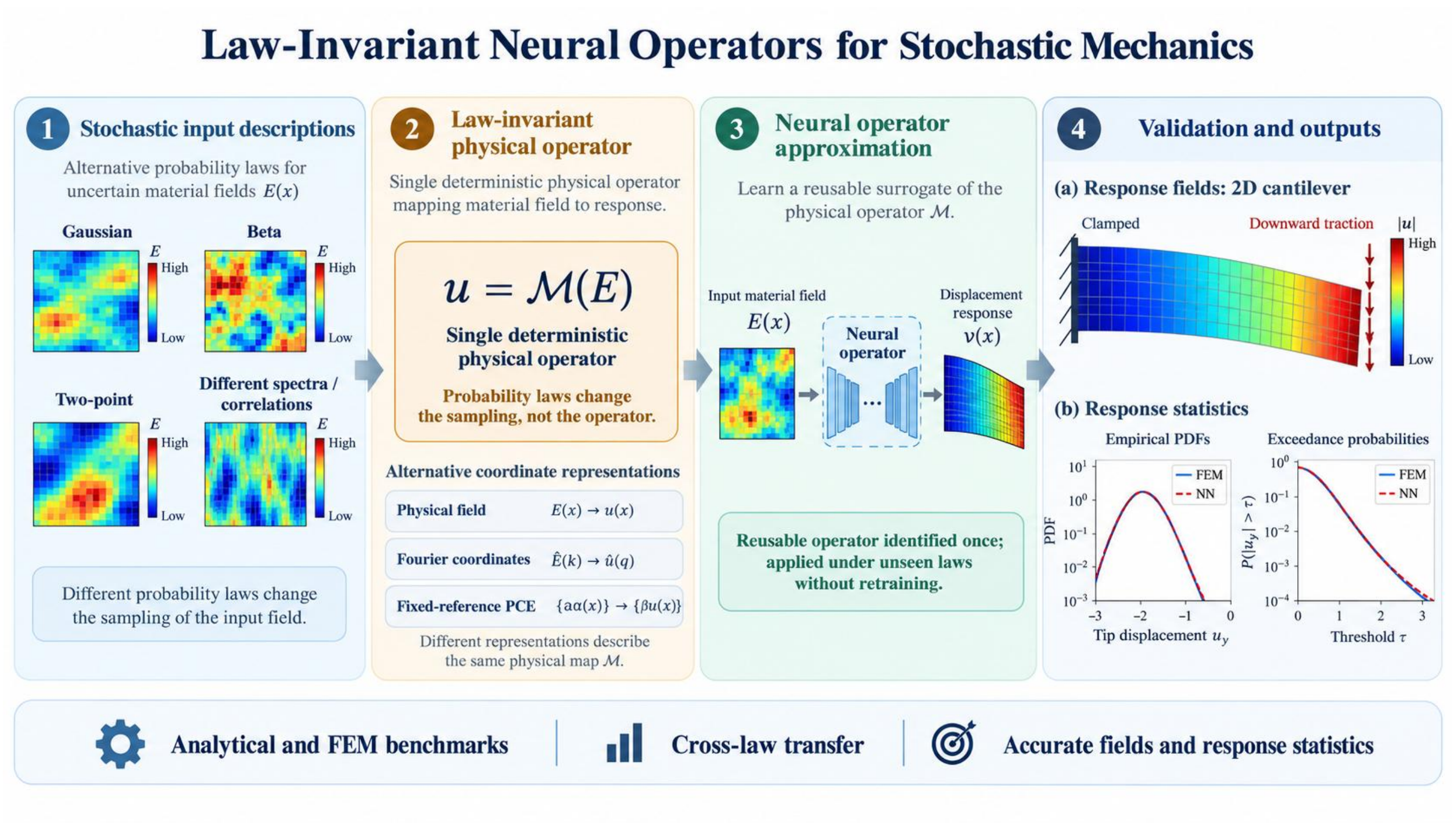

**Description.** The proposed framework separates the deterministic physical solution operator from the probability laws governing uncertain material fields. Physical-space, Fourier, and fixed-reference polynomial-chaos representations provide complementary descriptions of the same law-invariant operator. Neural approximations are evaluated through analytical and finite-element benchmarks to investigate cross-law generalization and the prediction of stochastic response fields and statistics without retraining.

## 1. Introduction

Uncertainty in material properties, geometry, loads, boundary conditions and environmental actions is intrinsic to engineering prediction. When uncertain quantities vary spatially, they are naturally represented as random fields, and deterministic field equations become stochastic boundary-value problems. In structural mechanics, the resulting stochastic finite-element problem may require repeated solutions of large deterministic systems, often under many alternative assumptions for the marginal distribution, covariance structure, correlation length, or dependence model of the uncertain inputs. Monte Carlo simulation remains broadly applicable but can be prohibitively expensive when each deterministic solve is costly. Spectral stochastic finite-element methods and polynomial-chaos approaches were developed precisely to reduce this burden by representing the random input and response in a finite stochastic basis approximation [1–5].

A complementary line of work in structural stochastic mechanics seeks response functions that isolate the deterministic system from the stochastic spectrum. The variability response function (VRF) introduced by Shinozuka [6] is a prominent example. In this classical setting, response variance can be represented by an integral of a deterministic response function against the power spectral density of the input random field. This construction enables inexpensive sensitivity analysis with respect to correlation structure and was later extended to provide a clear interpretation of the wavelengths that control response variability of general stochastic systems [7, 8]. The recent review by Deodatis et al. [9] documents extensions of the VRF concept across one- and two-dimensional structures, multiple stochastic properties, large variability and nonlinear constitutive behavior. Nevertheless, a fundamental limitation remains: once the response becomes a nonlinear functional of a non-Gaussian random field, a single second-order spectrum and a single scalar response kernel cannot, in general, encode the full stochastic interaction.

In parallel, operator learning has emerged as a powerful framework for approximating mappings between function spaces. DeepONet introduced a branch–trunk architecture motivated by universal approximation of nonlinear operators [10]. The Fourier Neural Operator (FNO) and the general neural-operator framework parameterize nonlocal mappings efficiently, often using Fourier-domain kernels, and have demonstrated large speedups for repeated PDE solution [11,12]. Recent reviews emphasize that generalization beyond the training distribution, incorporation of known operator structure, and principled uncertainty quantification remain central open problems [13,14]. These issues are particularly relevant for stochastic mechanics, where a surrogate trained under one stochastic model may be asked to operate under a different marginal distribution, covariance model or correlation length. In this direction, recent work has brought classical stochastic mechanics ideas into operator learning. Neural Chaos replaces fixed polynomial-chaos basis functions with data-driven orthogonal neural basis functions [15]. Sharma, Novák and Shields formulate polynomial chaos itself as an operator-learning method [16]. Shi et al. derive Wiener-chaos neural operators for SDEs and SPDEs, making the separation between stochastic forcing and deterministic chaos-coefficient dynamics explicit [17]. In a different direction, Mohseni, Duffield and Wong use Volterra expansions and Fourier-domain constructions to obtain interpretable higher-order convolutional representations of nonlinear operators [18]. Gopakumar et al. similarly argue that explicit decomposition into physical operators can improve extrapolation to new regimes [19]. These developments have sought to improve the generalization capabilities of learned operators across different input realizations, parameter configurations, and stochastic environments. Despite these advances, the influence of stochastic training distributions, spatial correlation structures, spectral excitation characteristics, and nonlinear mode interactions on the accuracy, efficiency, and generalization capabilities of learned operators remain insufficiently understood and constitute an open issue.

The central premise of this work is illustrated in Figure 1: A fixed physical system defines a deterministic solution operator independently of the probability laws assigned to its uncertain inputs. Different stochastic laws change how admissible inputs are sampled and hence the resulting response statistics, but not the underlying physical mapping. The present paper formalizes this law invariance and examines the same physical transfer through its finite-element matrix realization and through complete, fixed coordinate representations, including physical-field, Fourier, and fixed-reference polynomial-chaos descriptions. It is demonstrated that exact law invariance is a property of the physical operator. The accuracy of its learned approximation under unseen laws additionally depends on training coverage, representation, and neural architecture.

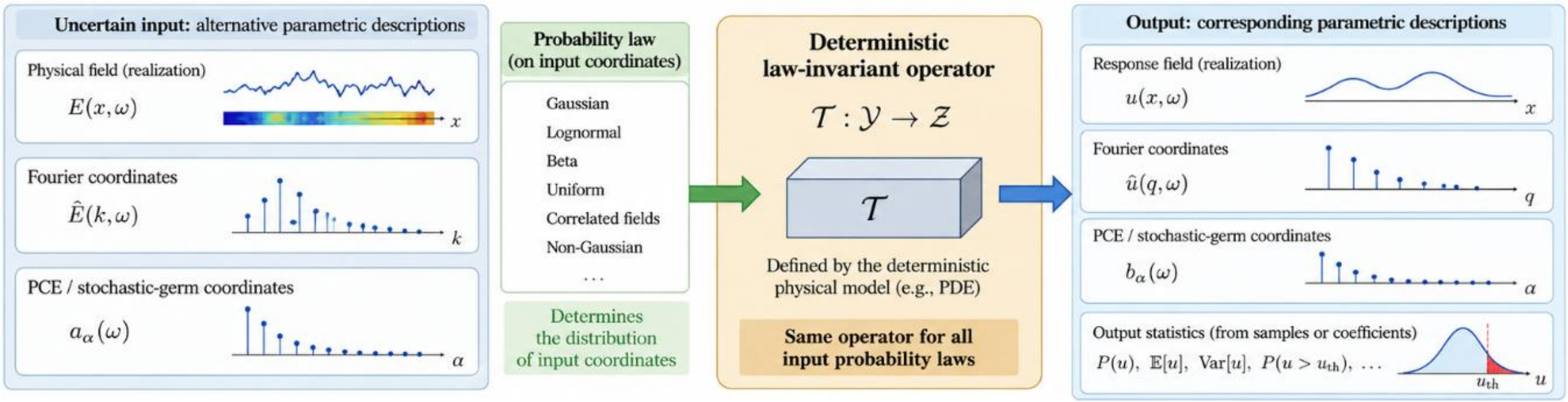

**Figure 1**. Conceptual framework for law-invariant uncertainty propagation. Different input probability laws determine how admissible material fields are sampled and induce distinct response distributions through the same deterministic physical operator. Physical-space, Fourier, and fixed-reference PCE descriptions provide complementary representations; in the finite-element setting, the physical map is realized by the inverse-stiffness action on the applied load.

The proposed framework, inspired by the fundamental philosophy of classical Variability Response Functions (VRFs), generalizes and completes the separation of the physics-based deterministic structural response characteristics, from the statistical description of uncertain inputs. Whereas classical VRFs enable the reuse of deterministic response information for evaluating second-order statistics under different stochastic spectra, the present approach extends this separation principle to the complete input–response mapping. In Fourier coordinates, nonlinear dependence on uncertain inputs introduces higher-order cross-frequency interactions, while the complementary fixed-reference polynomial-chaos (PCE) formulation describes coupling among stochastic modes. Both provide representations of the same underlying law-invariant physical operator. In a classical finite-element setting, this principle finds a direct algebraic interpretation through the deterministic assembly and inverse action of realization-dependent stiffness matrices, independently of their probability laws. Building on these complementary formulations, the numerical investigations examine cross-law generalization and the influence of stochastic identification measures, excitation characteristics, and neural architectures on response accuracy and statistical predictions, demonstrating that the law invariant operator (LIO) does not guarantee comparable generalization accuracy of its neural approximation across different stochastic laws.

The numerical program comprises five study configurations. A preliminary quadratic Fourier example examines transfer with analytically supplied nonlinear interaction features. A physical-coordinate DeepONet for a flexural cantilever and a full Fourier-coordinate surrogate for an axially loaded bar evaluate finite-element inverse-action reuse under changing modulus laws, including artificially bounded white-noise field identification. A statically indeterminate clamped–clamped beam then provides the principal comparison of broadband, correlated, monochromatic, and two-tone excitation measures in Fourier and fixed-reference PCE coordinates, including response PDFs and exceedance probabilities. Finally, a two-dimensional plane-stress finite-element benchmark examines supervised cross-law transfer in spatially heterogeneous elasticity. Together these cases assess the practical identification and reuse of a common deterministic physical operator across alternative stochastic descriptions.

## 2. Background and positioning of the proposed framework

### *2.1 Stochastic finite elements and spectral uncertainty propagation*

Spectral stochastic finite-element methods (SFEM) represent uncertain material fields and structural responses through finite stochastic bases. The classical work of Ghanem and Spanos [1,2] combined Karhunen–Loève random-field expansions with polynomial chaos. Generalized polynomial chaos extended this approach to broader probability families [3,4], while stochastic Galerkin formulations established variational frameworks for differential equations with random coefficients [5]. Arbitrary polynomial chaos further accommodates distributions specified by moments, histograms, or data [21]. Such methods can substantially reduce the cost of uncertainty propagation for a given stochastic model; however, changes in input laws or spatial dependence may require revised stochastic representations or additional computations. This becomes particularly relevant when several plausible statistical characterizations of the same uncertain material properties must be assessed.

When input statistics are uncertain, repeated analyses across alternative marginal distributions, coefficients of variation, correlation lengths, and power spectral densities (PSDs) can remain computationally demanding. Classical VRFs pursue a complementary separation: a deterministic structural response characterization is combined with the input PSD to evaluate response variance [6-8]. This allows the deterministic characterization to be reused for different spectra and reveals the wavelengths controlling response variability. For suitable problem classes, VRF formulations can also support response-variability bounds using admissible spectral information. The underlying VRF philosophy motivates the present work. For general nonlinear responses, particularly under non-Gaussian inputs, however, second-order spectra and a single scalar response function cannot fully represent higher-order dependencies and interactions among spatial frequencies [8, 9].

### 2.2 Neural operator learning and stochastic identification measures

Neural operators approximate mappings between input and response functions, offering a route to efficient repeated PDE solutions. DeepONet uses branch and trunk networks to represent nonlinear operators [10], whereas the Fourier Neural Operator (FNO) employs learned spectral operations [11]. More general operator-learning formulations address function-space approximation and discretization transfer [12–14]. Although the exact physical solution map is deterministic, a neural approximation is trained on a finite ensemble of input fields. Its accuracy may therefore change with the marginal law, spatial correlation, amplitude, or spectral content of the test inputs.

Recent approaches connect stochastic representations with operator learning. Neural Chaos constructs learned orthogonal stochastic bases [15]; polynomial-chaos operator learning uses structured stochastic expansions [16]; and Wiener-chaos neural operators exploit chaos-coefficient dynamics for stochastic differential equations [17]. Higher-order Fourier–Volterra representations describe nonlinear cross-frequency interactions [18], while physics-aware decompositions incorporate known operator structure [19]. Kernel-based methods have also been proposed as competitive alternatives for learning operators between function spaces [20]. These developments provide valuable representation and approximation tools but do not, by themselves, establish reliable transfer to unseen stochastic input laws.

A key distinction is between the fixed stochastic space or coordinate basis used to represent an input, the identification measure used to train its neural approximation, and the physical probability law used for subsequent uncertainty propagation. The pertinent question is not whether deterministic physical solution operators depend on input probability laws—they do not—but how their approximations can be identified accurately and reused when those laws change. The present framework connects physical-field, Fourier, and fixed-reference polynomial-chaos descriptions with the inverse-stiffness action of finite-element equilibrium. Numerical studies then examine correlated, bounded-white, monochromatic, and multifrequency identification ensembles to assess cross-law transfer and response-statistics fidelity, without presuming that any single training measure is universally optimal. This distinction is especially relevant when the quantities of interest include full response distributions and small exceedance probabilities, rather than only displacement-field approximation errors.

## 3. LIOs: Law-invariant deterministic transfer operators

### 3.1 Deterministic solution operator and stochastic excitation

Let $X$ be an admissible input space and $Z$ a response space. A deterministic parameter vector $d$ fixes the geometry, deterministic constitutive quantities, boundary conditions, deterministic forcing and the discretization/model choices. The governing equations define a deterministic solution operator

$$\mathcal{M}_d : X \to Z, \quad U = \mathcal{M}_d[Y] \tag{1}$$

When $Y$ is an $X$-valued random element defined on a probability space $(\Omega, F, P)$, the response $U$ is a corresponding $Z$-valued random event. The probability law $P(Y)$ specifies how the admissible input states are sampled; it does not enter the definition of the deterministic map $\boldsymbol{\mathcal{M}}(d)$ itself. Uncertainty propagation is consequently the push-forward of the input law through a fixed physical map

$$P_U = (\mathcal{M}_d)_\# P_Y \tag{2}$$

The governing physics equations define completely the deterministic solution operator $\mathcal{M}(d)$. However, the learning task here is to construct a computationally efficient approximation of its realization-to-response map. For an input random field $Y$ with law $P(Y)$, this map produces the stochastic response $U = \mathcal{M}_d[Y]$, whose distribution depends on $P(Y)$. Cross-law transfer therefore tests whether the learned approximation remains accurate for inputs drawn from different laws. This is distinct from learning a map from an input law directly to response statistics, which would itself depend on the law.

### *3.2 Finite-element specialization: the realization-dependent inverse stiffness operator*

Consider a linear FEM problem with uncertain material properties and let $E$ denote the discretized material field and f the assembled external load vector. For each admissible $E(\omega)$ realization, the equilibrium equations are

$$K(E(\omega); d)\, u(\omega) = f \tag{3}$$

For linear elastic finite element systems the assembled stiffness matrix depends linearly on the element moduli, while its inverse generally introduces a nonlinear dependence on these parameters. Assume that $K(E)$ is nonsingular for every $E$ in the admissible set. Define the realization-dependent inverse stiffness, or discrete solution/resolvent operator,

$$\mathcal{R}_E := K(E(\omega))^{-1} \tag{4}$$

Then, for a fixed load $f_0$, the stochastic finite-element response is simply

$$u(\omega) = \mathcal{R}_E f_0 = K^{-1} f_0 \tag{5}$$

The notation $K^{-1}$ is therefore shorthand for $K(E(\omega))^{-1}$. Although the stiffness matrix becomes random when the material field is stochastic, its inverse action is deterministic for each moduli realization and independent of the probability law used to generate it. Thus, the fixed-load material-to-response mapping provides a direct finite-element representation of the law-invariant physical operator. Importantly, even when the stiffness matrix depends linearly on the material properties, the inverse-stiffness mapping is generally nonlinear.

$$(K + \Delta K)^{-1} - K^{-1} = -K^{-1} \Delta K (K + \Delta K)^{-1} \tag{6}$$

This follows from the resolvent identity which shows why perturbations of $E$ generate nonlinear response [23]. If $K(E)$ is differentiable, the first directional derivative of the fixed-load solution is

$$Du[E](\delta E) = -K(E)^{-1} DK[E](\delta E) u(E) \tag{7}$$

If $K$ is linear in $E$, $D^2K=0$ and higher derivatives arise entirely from higher order inverse-stiffness operations. For example, the mixed second derivative is

$$D^2u[E](\delta E_1, \delta E_2) = K^{-1} DK(\delta E_1) K^{-1} DK(\delta E_2) u + K^{-1} DK(\delta E_2) K^{-1} DK(\delta E_1) u. \tag{8}$$

These derivatives characterize the nonlinear sensitivity of structural response to material-property variations and provide the physical-space counterparts of the higher-order transfer kernels subsequently developed in Fourier coordinates.

### *3.3 The learned deterministic solution operator and its representations*

Neural networks can successfully approximate material-to-response mappings, such as those described in Section 3.2. Such parametric solution approximations are well established, ranging from conventional feed-forward neural networks operating on fixed discretizations to more general operator-learning architectures, including DeepONet and Fourier Neural Operators (FNOs) [12–14]. Under prescribed loading, the learned mapping approximates the action of the inverse stiffness matrix on the applied load rather than the complete inverse matrix. When both material properties and loading vary, the formulation extends naturally to a parametric solution operator.

In conventional physical-space implementations, learned mappings associate material-property fields with corresponding response fields. Alternatively, a Fourier-coordinate formulation expresses the mapping directly through the Fourier coefficients of the input and output fields, consistent with related spectral-coefficient operator-learning approaches [22]. This representation provides a basis for characterizing spatial-frequency transfer and nonlinear cross-frequency interactions. Unlike conventional FNOs, which employ Fourier operations within their network layers, the formulation considered here explicitly uses Fourier coefficients as network inputs and outputs. Similarly, a fixed-reference polynomial-chaos (PCE) formulation induces a deterministic mapping between stochastic expansion coefficients, describing nonlinear coupling among stochastic modes. Complete representations preserve the underlying physical mapping, whereas truncation introduces additional approximation errors.

These formulations describe the same deterministic physical solution operator but differ in their coordinate representations, neural architectures, and approximation efficiency. Their learned approximations may consequently exhibit different accuracies across stochastic input laws, depending on training coverage, representation properties, and model capacity. The present work does not seek to introduce these established approximation strategies, but rather to investigate their reuse under different stochastic laws and the influence of identification measures on cross-law generalization.

### 3.4 General coordinate-conjugacy formulation

Let $\mathcal{M}(d): X \to Z$ denote the deterministic solution operator for fixed physical parameters $d$. Consider fixed, complete, and invertible input and response coordinate transformations $C_X: X \to X_C$ and $C_Z: Z \to Z_C$ . The corresponding representation of the physical solution operator is

$$\mathcal{T}_C^{(d)} = C_Z \circ \mathcal{M}(d) \circ C_X^{-1} \tag{9}$$

This expression defines an exact change of coordinates: the original mapping can be recovered as $\mathcal{M}(d) = C_Z^{-1} \circ \mathcal{T}_C^{(d)} \circ C_X$. Since the physical operator and the fixed coordinate transformations contain no input probability-law parameter, the transformed operator remains independent of the law used to sample its inputs.

For complete discrete Fourier transforms $\mathcal{F}_E$ and $\mathcal{F}_U$ of the material and response fields, respectively, the general expression specializes to

$$\hat{u} = \mathcal{T}_F^{(d)}(\hat{E}), \qquad \mathcal{T}_F^{(d)} = \mathcal{F}_U \circ \mathcal{M}(d) \circ \mathcal{F}_E^{-1} \tag{10}$$

Thus, the physical-space and full Fourier-coordinate formulations express the same deterministic material-to-response mapping. Any differences between independently trained neural approximations arise from the chosen representation and learning procedure, rather than a change in the underlying physics. This exact coordinate equivalence requires complete, invertible transformations; spectral truncation introduces an additional representation error. The fixed-reference PCE coefficient formulation is considered separately in Section 5, where synthesis and projection need not constitute an invertible coordinate transformation after truncation.

The corresponding stochastic response laws are consistently transformed under these coordinate changes, since applying the physical solution operator before or after the associated coordinate transformations yields equivalent probability distributions.

### 3.5 Learning hypothesis and domain of validity

Previous sections 3.1-3.4 are exact statements about deterministic physical maps and their coordinate representations. Cross-law generalization of a finite neural approximation is not exact: it depends on training coverage, regularity, model capacity and the severity of the distribution shift. The hypotheses below are therefore evaluated empirically in Section 6, with conclusions restricted to the reported input families and numerical ranges.

**Hypothesis H1 — Cross-law generalization:** Let a neural LIO $\mathcal{N}_\theta$ approximate a fixed deterministic transfer *T(d)* using training samples drawn from a mixture of stochastic laws. If an unseen law generates inputs within a region of coordinate space sufficiently covered by training and *T(d)* is regular on that region, then the frozen approximation $\mathcal{N}_\theta$ should retain comparable accuracy without being given the law label. Generalization can fail when the unseen law places mass outside the training support, activates unresolved modes, or changes dependence structures not represented in the training set.

**Hypothesis H2 — Coordinate choice affects learnability, not exact physics:** Physical, Fourier and PCE coordinate maps are exactly equivalent when the transforms are invertible and complete. Their neural operator approximations need not be equivalent because sparsity, smoothness, mode coupling, truncation error, boundary effects and optimization conditioning differ between coordinate systems.

**Hypothesis H3 — FEM inverse-action reuse:** For a fixed deterministic finite-element model, a surrogate trained over a sufficiently rich set of admissible input material fields should approximate the same inverse-stiffness action $E \mapsto K(E)^{-1}f$ across unseen probability laws, provided those laws do not drive the system outside the sampled admissible domain.

With these hypotheses established, we next examine how different coordinate representations shape the learning problem and the conditions under which the resulting approximation can be validated.

## 4. Fourier-coordinate formulation

For a periodic or suitably discretized field, let $F$ denote the Fourier transform. Define $\hat{Y}=FY$ and $\hat{U}=FU$. The exact Fourier-coordinate transfer is written as

$$\hat{U} = T_d^F[\hat{Y}], \qquad T_d^F = F \circ \mathcal{M}_d \circ F^{-1} \tag{11}$$

This identity is independent of whether the Fourier coefficients are Gaussian, non-Gaussian, independent, dependent or generated by a translation field. The complete joint law of the coefficients controls stochastic statistics, while *T(F,d)* controls deterministic response transfer.

### *4.1 Linear limit and recovery of the VRF*

If the deterministic physical operator is linear, its Fourier representation is a linear kernel

$$\hat{U}(q) = \int H_d^{(1)}(q,k)\hat{Y}(k)\,dk \tag{12}$$

For a translation-invariant system, the kernel is diagonal in frequency, *H1(d)*(q,k)=*H(d)*(k)δ(q−k), and therefore

$$\hat{U}(k) = H_d(k)\hat{Y}(k) \tag{13}$$

For a zero-mean stationary input with power spectral density *S(Y)*, the response variance becomes

$$\mathrm{Var}[U] = \int |H_d(k)|^2 S_Y(k)\,dk \tag{14}$$

The resulting variance expression recovers the classical spectral separation between a deterministic response function and the input power spectral density, with $|H(d)(k)|^2$ representing the VRF in this translation-invariant linear limit. In bounded finite-element domains, translation invariance generally does not hold, and the first-order Fourier transfer kernel $H^1(q, k)$ is generally non-diagonal. Consequently, different input frequencies may contribute to each output frequency even for linear operators. This distinction is particularly relevant for structures with boundary constraints, localized loads, interfaces, or spatially varying deterministic coefficients.

### *4.2 Nonlinear cross-frequency transfer*

Assume ***M(d)*** is sufficiently smooth around a reference input $\bar{Y}$. A functional Taylor/Volterra expansion may be written formally as

$$U = U_0 + \sum_{n=1}^{\infty} \mathcal{M}_d^{(n)}[\delta Y, \dots, \delta Y], \qquad \delta Y = Y - \bar{Y} \tag{15}$$

After Fourier transformation, each nth-order term produces an n-frequency transfer kernel

$$\hat{U}(q) = \hat{U}_0(q) + \sum_{n=1}^{\infty} \int \cdots \int H_d^{(n)}(q; k_1, \dots, k_n) \prod_{j=1}^{n} \hat{Y}(k_j)\, dk_1 \cdots dk_n \tag{16}$$

For translation-invariant local polynomial nonlinearities, frequency-selection rules such as $q=k1+\cdots+kn$ arise. For bounded or nonhomogeneous systems, the kernels generally retain broader q–k coupling. The crucial point is that all $H^n(d)$ are properties of deterministic physics; stochastic information enters through the joint moments or laws of the random Fourier coefficients.

This observation resolves the difficulty encountered by generalized effective-VRF formulations [9]: A nonlinear response cannot generally be written as a single stochastic-independent scalar kernel multiplying $S(Y)(k)$, yet deterministic–stochastic separation remains valid at the operator level. Higher-order kernels replace the scalar VRF.

## 5. Polynomial-chaos coordinate formulation

To preserve a common coordinate representation across changing stochastic laws, the PCE formulation is posed on a fixed reference probability space (Γ, ρ). Let $\Psi_\alpha(\xi)$ be an orthogonal basis with $E_\rho[\Psi_\alpha\Psi_\beta]=\gamma_\alpha\delta_{\alpha\beta}$. An uncertain input field represented in this reference coordinates has the following expansion

$$Y(x,\xi)=\sum_\alpha y_\alpha(x)\Psi_\alpha(\xi),\qquad U(x,\xi)=\sum_\beta u_\beta(x)\Psi_\beta(\xi) \tag{17}$$

Different physical stochastic laws can be represented through different coefficient sets, or through deterministic transports from the same reference configuration. This fixed-reference requirement is important: if a different law-adapted basis is chosen for every marginal distribution, the coordinates themselves change and direct coefficient-level law invariance becomes ambiguous.

Projection of the deterministic response onto the fixed basis gives

$$u_\beta(x)=\frac{1}{\gamma_\beta}\int_\Gamma \mathcal{M}_d\left[\sum_\alpha y_\alpha(x)\Psi_\alpha(\xi)\right](x)\Psi_\beta(\xi)\rho(\xi)\,d\xi \tag{18}$$

Together, the input expansion in eq(17), the deterministic response $\mathcal{M}(d)$ and the projection in eq(18) define a deterministic coefficient map $T_{PC,d}:\{y_\alpha\}\to\{u_\beta\}$. For a fixed complete input coefficient set, this map is independent of the external probability-law label, provided the physical operator, reference probability space, and basis remain fixed. A change in law may alter the coefficient set or the transport used to generate the physical field. If each law instead uses its own basis, the coefficient sets must first be mapped to a common basis for comparison.

### *5.1 Linear and nonlinear stochastic-mode coupling*

For linear physics, $\mathcal{M}(d)$ commutes with the stochastic expansion and the modes decouple:

$$u_\alpha=\mathcal{M}_d[y_\alpha]\qquad(\text{linear } \mathcal{M}_d) \tag{19}$$

For nonlinear physics, modes couple. If $\mathcal{M}(d)$ is expanded into multilinear terms, the quadratic contribution has the form

$$u_\beta=\mathcal{M}_d^{(1)}[y_\beta]+\sum_{\alpha,\gamma} C_{\beta\alpha\gamma}\mathcal{M}_d^{(2)}[y_\alpha,y_\gamma]+\cdots,\qquad C_{\beta\alpha\gamma}=\frac{\mathbb{E}_\rho[\Psi_\beta\Psi_\alpha\Psi_\gamma]}{\gamma_\beta} \tag{20}$$

Hence Fourier and PCE representations reveal complementary structures: spatial nonlinearity appears as frequency coupling, while stochastic nonlinearity appears as chaos-mode coupling. The neural operator may be viewed as a compact approximation to the corresponding high-order transfer hierarchy. Once the output PCE coefficients are available, response statistics may be obtained algebraically.

## 6. LIO identification and cross-law generalization

Sections 3–5 describe a deterministic physical solution operator and its finite-element, Fourier, and fixed-reference polynomial-chaos (PCE) formulations. The remaining question is how accurately a finite neural approximation of this operator can be identified from sampled input–response pairs and reused under stochastic laws not represented during training. In this setting, the probability law governing the uncertain physical input must be distinguished from the identification measure used to construct the training ensemble.

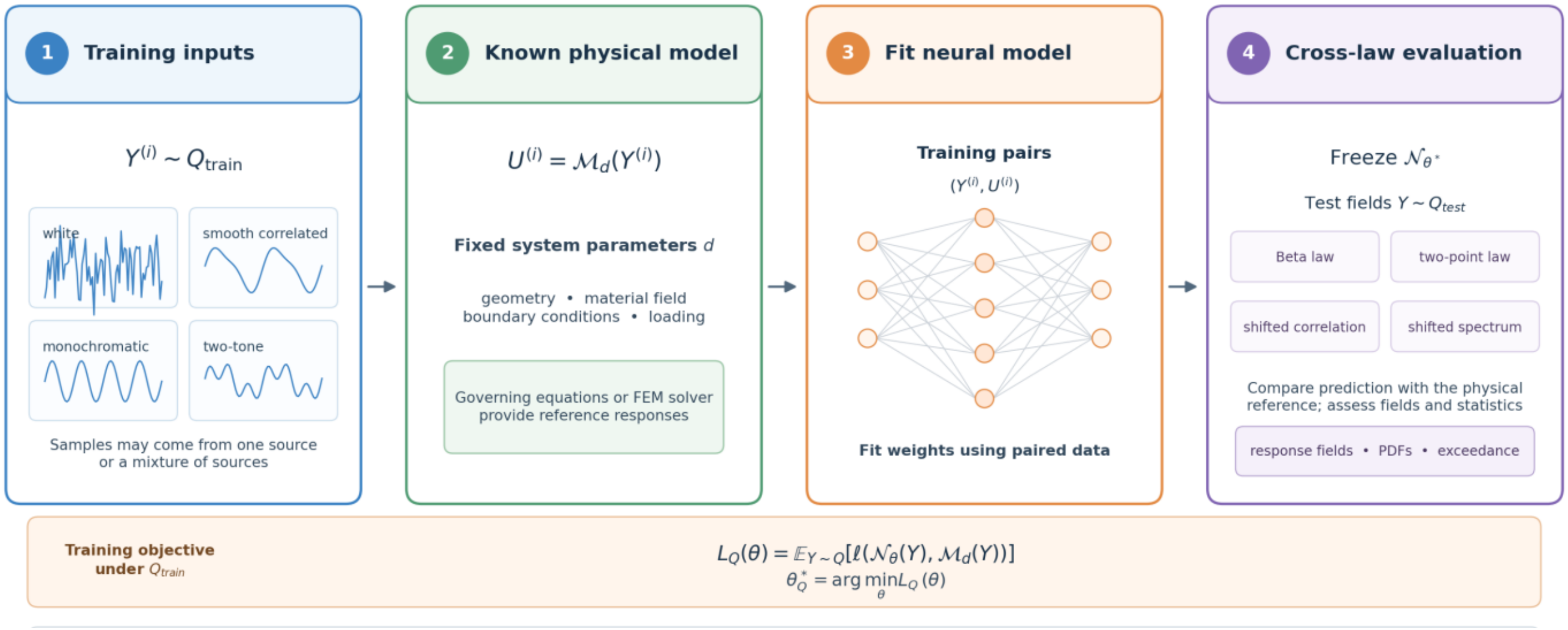

**Figure 2.** Learning framework for LIO identification and cross-law generalization.

Figure 2 summarizes the learning framework: Training samples drawn from a selected identification measure are passed through the deterministic physical operator to generate response data, from which a neural approximation is constructed. The trained surrogate is then evaluated without retraining under different stochastic input laws, including held-out distributions and out-of-distribution conditions. The figure also distinguishes exact law invariance from the dependence of surrogate accuracy on training coverage, coordinate representation, and architecture.

### *6.1 Identification measures and learning objectives*

Let $\mathcal{M}^{(d)}: \mathcal{X} \to \mathcal{Z}$ be the deterministic solution operator for the fixed system parameters $d$, and let $Q$ denote a probability measure on the admissible input fields $Y \in \mathcal{X}$ used exclusively for operator identification. A neural approximation $\mathcal{N}_\theta$ is obtained by minimizing an expected discrepancy between the predicted and reference responses:

$$\mathcal{L}_Q(\theta) = \mathbb{E}_{Y\sim Q}\left[\ell\left(\mathcal{N}_\theta(Y), \mathcal{M}^{(d)}(Y)\right)\right], \qquad \theta_Q^* \in \arg\min_\theta \mathcal{L}_Q(\theta). \tag{21}$$

Here $\ell$ is the chosen training loss; in the numerical studies it is evaluated from deterministic analytical or finite-element response solutions. For a prescribed structural load, the neural model $\mathcal{N}_\theta(Y)$ approximates the corresponding material-to-displacement solution map, rather than the entire inverse-stiffness matrix.

In this process, the identification measure need not coincide with any particular physical uncertainty model. It may involve an artificial, positive bounded white-field ensemble, spatially correlated fields, or deliberately structured monochromatic and multifrequency excitations. Multiple sources can also be combined through a mixture

$$Q = \sum_{j=1}^{J} \alpha_j Q_j, \qquad \alpha_j \geq 0, \qquad \sum_{j=1}^{J} \alpha_j = 1, \tag{22}$$

where the weights control the relative frequency of different input families during training. The central identification issue is therefore not whether the exact physical operator changes with $Q$—it does not—but how the sample coverage induced by $Q$ affects the accuracy of its finite neural approximation.

For a set of complete input and output representations $C_X, C_Z$, a network trained in transformed coordinates may be interpreted in physical space through

$$\mathcal{N}_\theta^{\text{phys}} = C_Z^{-1} \circ \mathcal{N}_\theta^{(C)} \circ C_X. \tag{23}$$

This interpretation applies directly to complete, invertible coordinate transformations, including the full discrete Fourier transform. A complete PCE representation on the fixed reference probability space also represents the same physical input–response map. A finite PCE truncation, however, introduces representation error and is not generally an invertible coordinate transformation. In Fourier-based field generation, an underlying Gaussian process may be transformed to produce fields with different marginal distributions, whereas in PCE the reference measure defines the orthogonal polynomial basis. In either case, the reference measure serves a different purpose from the identification measure, which governs the physical input fields sampled for neural training. Truncation of either the Fourier or PCE expansion introduces representation error and generally prevents exact equivalence with the original physical-space operator.

### <u>*6.2 Cross-law transfer and statistical fidelity*</u>

After identification under $Q$, the network parameters are frozen and performance is evaluated under a distinct stochastic input law $P$. A realization-wise relative response-field error, averaged over the test law, may be written as

$$\mathcal{E}_P\left(\theta_Q^*\right) = \mathbb{E}_{Y\sim P}\left[\frac{\left\|\mathcal{N}_{\theta_Q^*}(Y) - \mathcal{M}^{(d)}(Y)\right\|_2}{\left\|\mathcal{M}^{(d)}(Y)\right\|_2}\right], \tag{24}$$

for nonzero reference response fields, with the expectation estimated from an independent test ensemble. Cross-law transfer refers to the ability of a trained neural approximation to predict responses under previously unseen input probability laws without retraining. More demanding out-of-distribution (OOD) tests additionally modify spatial correlation, variability level, or other input characteristics beyond the training regime. These tests provide empirical evidence over specified ensembles; they do not establish uniform approximation accuracy for arbitrary stochastic laws. Prediction of stochastic response statistics is a related but distinct requirement. For a scalar response functional $g$, such as displacement at a selected point, the exact response is $R = g\left(\mathcal{M}^{(d)}(Y)\right)$, whereas the surrogate gives $\hat{R} = g\left(\mathcal{N}_{\theta_Q^*}(Y)\right)$. Of particular interest are the mean, standard deviation, response probability density, and exceedance probability

$$p_P(r_0) = \mathbb{P}_{Y\sim P}\left[g\left(\mathcal{M}^{(d)}(Y)\right) > r_0\right], \qquad \hat{p}_P(r_0) = \mathbb{P}_{Y\sim P}\left[g\left(\mathcal{N}_{\theta_Q^*}(Y)\right) > r_0\right]. \tag{25}$$

Even small response-field errors can alter exceedance classification for samples near the threshold $r_0$, particularly when the target probability is small. Accordingly, the numerical tests report field errors separately from response-distribution and tail-probability accuracy. Empirical exceedance estimates are also subject to finite-sample uncertainty, independently of surrogate error.

### <u>*6.3 Learning questions and scope of the numerical assessment*</u>

The following numerical investigation addresses three related questions. First, cross-law reuse asks whether one frozen approximation of a fixed physical system retains useful accuracy under input distributions absent from the identification dataset. Second, identification-measure design asks how broadband, correlated, monochromatic, and multifrequency training sources affect the approximation of nonlinear response dependence and the transfer to other stochastic input families. Third, representation and statistical fidelity asks how physical, Fourier, and fixed-reference PCE parameterizations and the corresponding neural architectures affect response-field accuracy and the reliability of inferred response statistics.

These are empirical learning questions rather than consequences of exact law invariance. Performance depends jointly on the admissible input domain, the coverage and size of the identification sample, the chosen representation, network capacity, and optimization. In particular, the different numerical configurations employ different discretizations, stochastic parameterizations, and architectures; comparisons between them must therefore not be interpreted as controlled rankings of representation classes. The analytical and finite-element experiments in Section 7 are designed to characterize these dependencies within the reported stochastic families and computational settings.

## 7. Numerical validation

The numerical program examines both exact operator mappings and coordinate identities, as well as the empirical behavior of learned neural approximations. It contains a controlled preliminary Fourier-learning experiment, two complementary finite-element cross-law benchmarks, a statically indeterminate clamped–clamped beam study, and a two-dimensional plane-stress FEM study. The clamped–clamped beam study examines how excitation measures affect Fourier and fixed-reference PCE approximation, whether learned models reproduce displacement PDFs and low exceedance probabilities, and how field and tail errors respond to increased training data and network capacity. The two-dimensional FEM study extends the cross-law assessment to supervised physical-coordinate learning and response-tail prediction. Unless otherwise stated, the field metric is the mean relative Euclidean response error computed for each realization. The metric is defined in eq(26).

$$e_{field} = \frac{100}{N} \Sigma_{i=1}^{N} ||u_i^{NO} - u_i^{ref}||_2 / ||u_i^{ref}||_2 \tag{26}$$

where $u_i^{NO}$ and $u_i^{ref}$ are the predicted and reference response vectors for realization $i$. This metric measures realization-wise response accuracy and is distinct from the error in the ensemble mean or standard-deviation field. Those moment errors are introduced separately when ensemble statistics are evaluated. All Fourier modes are retained in the reported Fourier learners, whereas finite PCE expansions have a separate stochastic-basis truncation error. Table 1 summarizes the five configurations, including the two-dimensional plane-stress FEM study introduced in Section 7.4.

| Case | System and representation | Scientific question |
|---|---|---|
| I | Quadratic map; structured Fourier neural transfer | Does a learned transfer reuse known nonlinear interaction features across an unseen law? |
| II | 50-element Euler–Bernoulli cantilever; physical-space DeepONet | Does a surrogate of the FEM inverse-stiffness action transfer between modulus laws? |
| III | 16-element axial-bar FEM; exact and learned Fourier coordinates | Can a Fourier map be learned across laws, including with an artificial iid identification measure? |
| IV | Analytical uniformly loaded cantilever; Fourier and fixed-germ PCE | Which combinations of white, smooth, single-frequency and two-frequency training fields identify the nonlinear operator efficiently? |
| V | 2D plane-stress cantilever FEM; supervised physical-coordinate MLP | Does the same fixed-load FEM solution map transfer to unseen modulus laws in 2D, and how accurately are response tails predicted? |

**Table 1**. Overview of the five numerical configurations. The quadratic study is a controlled preliminary test; the finite-element, clamped–clamped and two-dimensional studies provide the principal numerical evidence.

### *7.1 Case I: Preliminary experiment: structured quadratic Fourier transfer*

The nonlinear Fourier transfer mechanism of section 4 is first illustrated using the pointwise quadratic mapping $U_j = aY_j + bY_j^2$, where $a$ and $b$ are deterministic constants. Applying the discrete Fourier convolution theorem to the pointwise quadratic mapping yields the exact Fourier as

$$\hat{U}_q = a\hat{Y}_q + \frac{b}{N} \sum_{k=0}^{N-1} \hat{Y}_k \, \hat{Y}_{(q-k) \bmod N} \tag{27}$$

The quadratic term couples pairs of input frequencies whose indices sum to the output frequency modulo $N$. The corresponding transfer coefficients depend exclusively on the deterministic mapping, whereas the input probability law governs the statistical properties of the Fourier coefficients.

Let $Z$ be a zero mean and unit variance and length $L$=1 underlying Gaussian random field $Z(x, \omega)$ with covariance matrix $K_{jk} = C_Z(r_{jk})$, where $r_{jk}$ is the shortest periodic distance between grid points $x_j$ and $x_k$. The underlying covariance is periodized exponential $C_Z(r) = \frac{e^{-r/\ell} + e^{-(L-r)/\ell}}{1 + e^{-L/\ell}}$. This baseline test implements correlation length parameters $\ell/L = 0.10$, and a shifted spectrum with $\ell/L = 0.30$, as shown in Figure 3.

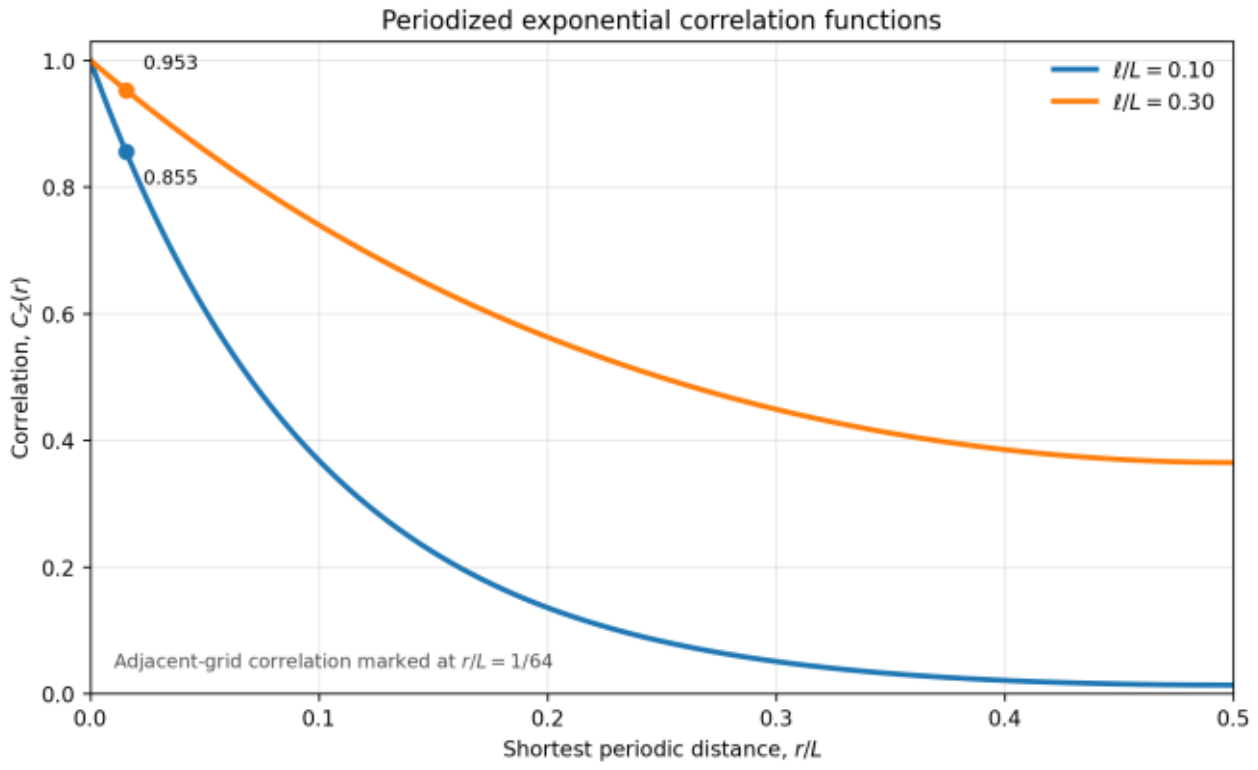

**Figure 3.** Correlation structure of the underlying Gaussian fields

Consider the input stochastic field $Y = Y(x, \omega)$ in eq(27). For the Gaussian marginal test, this is identical to the latent Gaussian field $Z$. For the other marginal law tests, component wise Gaussian-copula transforms map to unit-variance Laplace, centered-Gamma, or standardized Student-$t_5$ values with the following properties: Laplace scale $1/\sqrt{2}$; the centered-Gamma field is obtained from a Gamma distribution with shape 4 and scale 0.5 by shifting its mean by 2; and $\sqrt{(3/5)}$ times a standardized Student-$t_5$ variate as translation fields. These transforms preserve the latent Gaussian rank dependence but generally change the target PSD, so the stated spectrum refers to the underlying Gaussian. The training set contains 350 fields from each of the Gaussian, Laplace and centered-Gamma families, while validation contains 80 from each. Each test ensemble contains 500 fields. Student-$t_5$ is tested at both $\ell/L$ 0.10 and the shifted value $\ell/L = 0.30$.

For each field realization, a non-normalized DFT is applied to $Y$ and $Y^2$, giving $\hat{Y}(q)$ and the supplied quadratic feature $C_{2q} = F[Y^2](q)$. The target is $\widehat{U}(q) = \hat{Y}(q) + 0.35\, C_{2q}$. A shared 4→16→2 tanh MLP (114 parameters) maps the real and imaginary parts of those two coefficients to the real and imaginary parts of $\widehat{U}(q)$. Input and output channels are standardized using training data only. Training uses AdamW (learning rate 0.001, weight decay $10^{-6}$), batches of 1,024, a 300-epoch limit and 30-epoch validation patience; the fixed model seed is 12401 and the shuffle seed is 12501. The selected checkpoint is epoch 260. The metric is the per-realization relative Euclidean field error; by Parseval's identity it equals the full-spectrum relative Fourier error. One fixed training seed is used, and test-set summaries describe variability across the 500 realizations, not across training seeds.

Table 2 summarizes the Fourier-coordinate operator's test accuracy across the input marginal laws and latent correlation settings, using mean and percentile errors to show both typical performance and tail behavior. At the baseline spectrum, mean errors are 0.341% for Gaussian, 0.351% for Laplace, 0.341% for centered-Gamma, and 0.341% for the unseen standardized Student-$t_5$ fields; the 99th-percentile errors range from 0.787% to 0.918%. For Student-$t_5$ fields at the shifted spectrum, the mean error is 0.597%, the median is 0.307%, the 95th percentile is 1.425%, and the 99th percentile is 3.905%, with a maximum of 46.043%. Similar baseline results indicate accurate transfer across the tested marginal laws under the same latent dependence structure. Under the shifted spectrum, typical errors remain modest, but the larger upper-tail errors show sensitivity to the correlation-length change.

| Test ensemble (latent ℓ/L) | Mean error (%) | Median error (%) | 95th percentile (%) | 99th percentile (%) |
|---|---|---|---|---|
| Gaussian (seen; 0.10) | 0.341 | 0.294 | 0.618 | 0.912 |
| Laplace (seen; 0.10) | 0.351 | 0.296 | 0.688 | 0.918 |
| Centered Gamma (seen; 0.10) | 0.341 | 0.286 | 0.573 | 0.851 |
| Student-$t_5$ (unseen; 0.10) | 0.341 | 0.297 | 0.614 | 0.787 |
| Student-$t_5$ (shifted; 0.30) | 0.597 | 0.307 | 1.425 | 3.905 |

**Table 2**. Test-set relative physical-field: Mean, median, 95th percentile and 99th percentile computed over 500 realizations

Overall, this test supports cross-law transfer at the baseline correlation structure. However, it also shows that such transfer does not ensure robust accuracy under a shifted spectrum. The original 4–16–2 tanh neural operator evaluates the transfer of the supplied quadratic Fourier map rather than discovering its mode-coupling convolution. For the shifted latent correlation length $\ell/L$ = 0.30, the unseen Student-$t_5$ errors were 0.597%, 0.307%, 1.425%, and 3.905% for the mean, median, 95th percentile, and 99th percentile, respectively, with a maximum of 46.043%. Typical predictions remain accurate, whereas a small number of shifted-spectrum realizations produce the deterioration in the error tail. Because the quadratic Fourier feature $C_{2q}$ is supplied to the network, the target is affine in the four real input components. These errors therefore measure the tanh model's approximation and generalization error for shifted Fourier inputs, rather than its ability to identify nonlinear mode coupling.

As a structural control of the NO, a shared affine 4-to-2 map was fitted to the real and imaginary parts of $\hat{Y}(q)$ and $C_{2q}$ and applied to 5,000 shifted Student-$t_5$ test fields. This model produced relative zero errors for every realization, effectively eliminating the observed tail deterioration for this supplied-feature problem. This near-roundoff result confirms that the prescribed deterministic map transfers accurately when its affine structure is represented explicitly. It also shows that the tail error of the tanh model is a model-approximation error, rather than a failure of the underlying law-invariant physical map. The neural-operator architecture should therefore be selected according to the structure of the operator and the required accuracy. For this supplied-feature map, a shared affine model is better matched than a tanh network. Other problems involving genuine nonlinear interactions may require architecture capable of representing those interactions directly. No single neural-operator architecture is optimal for every problem. The two neural-operator structures are compared in Figure 4.

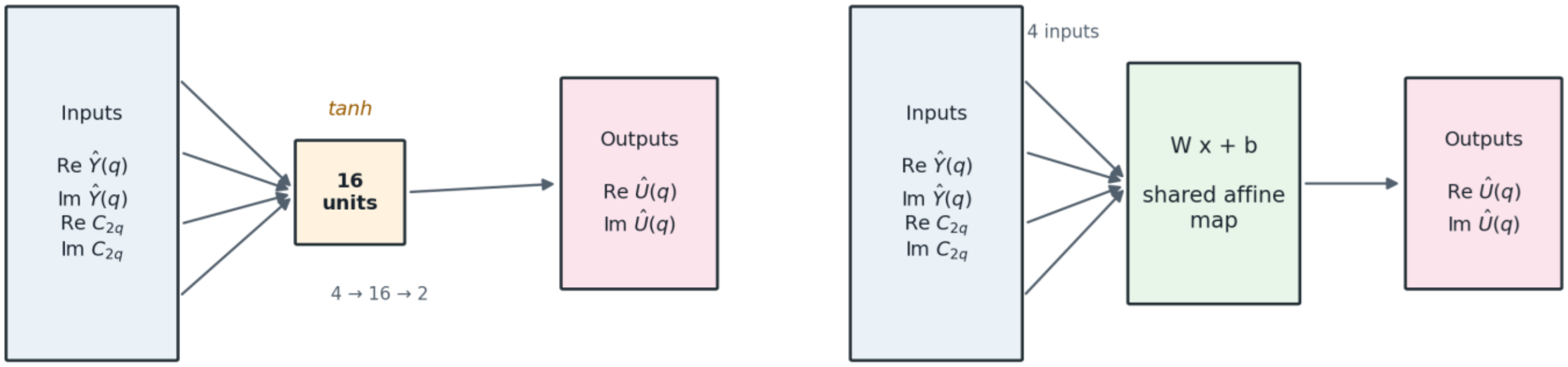

**Figure 4**. Fourier-coordinate neural-operator models: (a) The original model uses a shared 4 to 16 to2 tanh network, whereas the structure-matched model uses a shared affine 4 to 2 map.

*7.2 Complementary finite-element benchmarks: flexural and axial loading*

The two finite-element benchmarks of Figure 5 address complementary loading/deformation modes under stochastic Young's modulus. The two models use different meshes and neural architectures and were trained separately: (a) the bending cantilever is learned in physical coordinates, whereas the axially loaded bar is learned in Fourier coordinates.

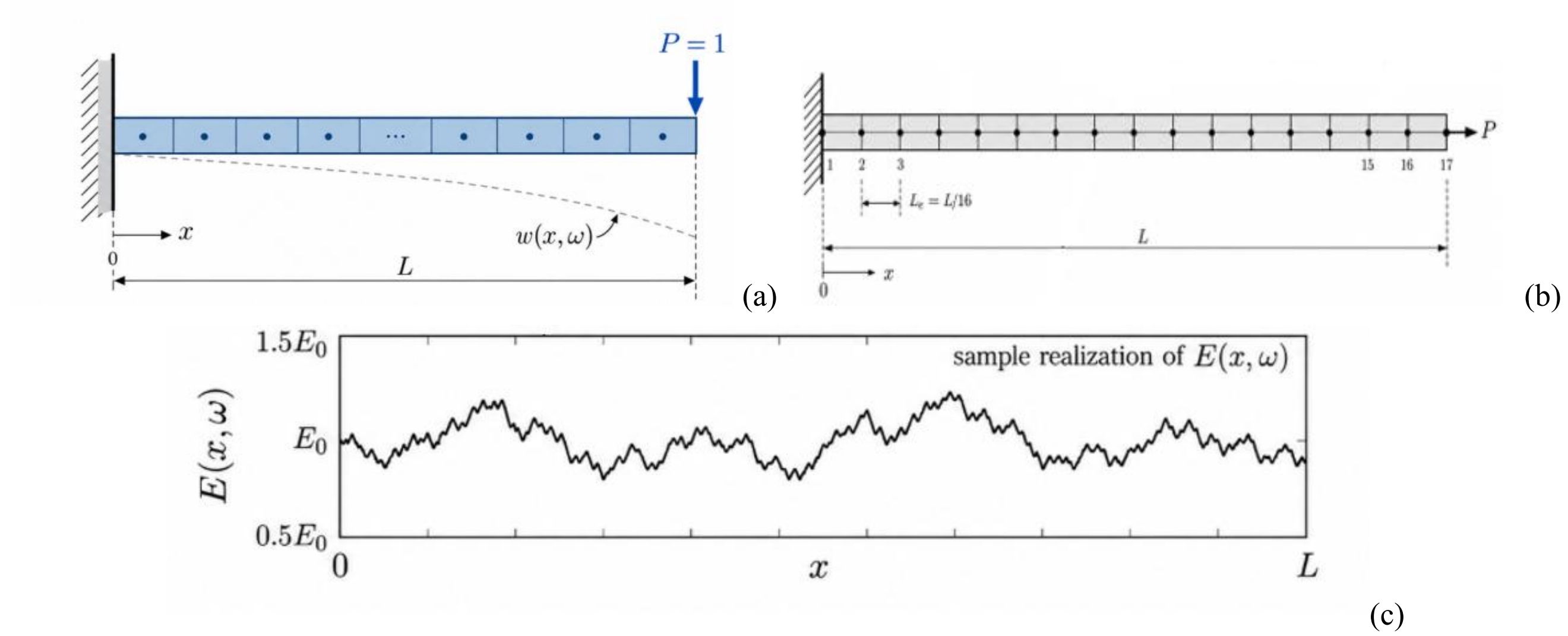

**Figure 5**. Complementary stochastic finite-element benchmarks: (a) a clamped Euler–Bernoulli cantilever under a transverse unit tip load and (b) an axially loaded 16-element bar, (c) Spatially varying modulus realization.

### *7.2.1 Case II: Flexural cantilever: DeepONet physical-coordinate model*

The unit-length cantilever beam (Figure 5(a)) was discretized into 50 two-node Euler–Bernoulli bending elements. The left end was clamped; a deterministic unit transverse point force was applied at the free end. Each element had a constant random modulus. The DeepONet branch network receives all 50 modulus values, while a spatial-coordinate trunk evaluated the transverse deflection field. The branch and trunk widths were 50–128–128–64 and 1–64–64–64, respectively, with 39,745 trainable parameters. Neither the marginal-law label nor stochastic covariance parameters were supplied to the network. The DeepONet architecture is illustrated in Figure 6.

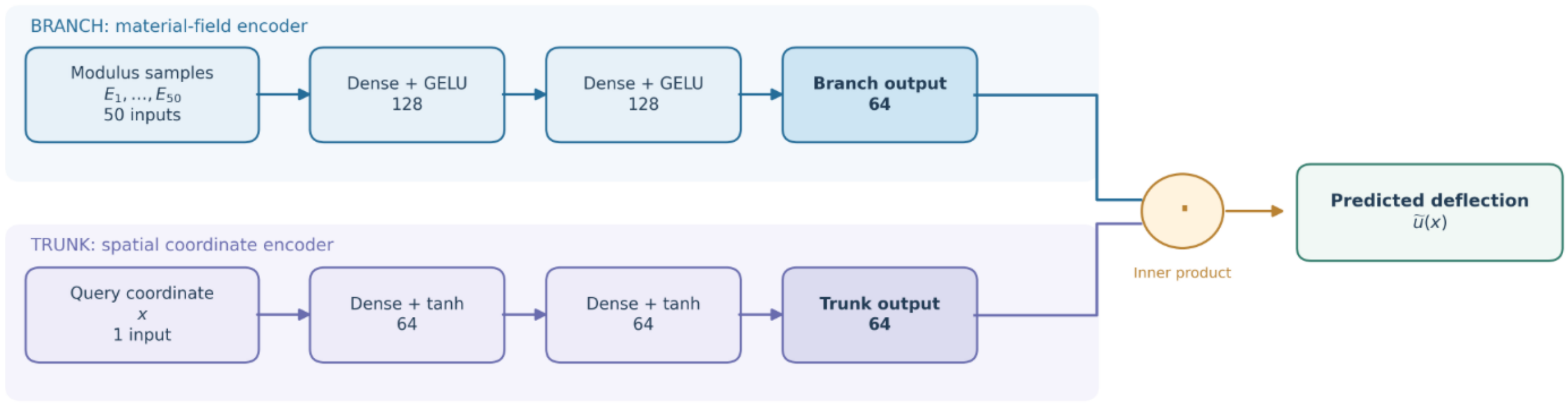

**Figure 6.** Physical-coordinate DeepONet for the 50-element flexural cantilever, consisting of a branch network (50–128–128–64) encoding the spatially varying Young's modulus and a trunk network (1–64–64–64) encoding the spatial evaluation coordinate.

The modulus stochastic fields were constructed from a unit-variance latent Gaussian vector $Z$ with exponential correlation at the element centroids:

$$R_{ij} = \exp\left(-\frac{|x_i - x_j|}{\ell_c}\right), \qquad Z = L_R \xi, \qquad L_R L_R^T = R, \quad \xi \sim \mathcal{N}(0, I). \tag{28}$$

Gaussian, lognormal and Gamma material marginals were constructed from this common latent field; the latter two employed as translational fields. Standardized Beta and two-point modulus laws were also held out. The Gaussian modulus was bounded (truncated) below to enforce positivity. Note that the specified exponential correlation belongs to the underlying field and nonlinear marginal transformations do not generally preserve the same physical correlation.

The training database comprised 900 independent FEM realizations, with 300 each from the Gaussian, lognormal and Gamma families. The latent correlation length was selected from $\ell_c/L \in \{0.05, 0.10, 0.20, 0.30\}$ and the modulus coefficient of variation from CoV ∈ {0.10, 0.20. A separate 180-realization set was used for selecting the network architecture, and checkpoint parameters, which were excluded from both training and final test-error evaluation. Beta and two-point marginals were reserved for unseen-law tests, including combined out-of-distribution cases with $\ell_c/L$ = 0.45 and CoV = 0.30.

Table 3 reports the prediction accuracy of the frozen DeepONet for seen and unseen Young's modulus distributions. The mean displacement-field errors for the unseen Beta and two-point laws are 0.558% and 0.726%, respectively, comparable to those obtained for the three training distributions (0.559–0.653%). Under the combined distributional and correlation/variability shifts, these errors increase to 1.149% and 1.780%, respectively. The corresponding 95th-percentile errors also increase, indicating greater sensitivity to out-of-distribution inputs. Nevertheless, the results demonstrate that a neural approximation of the deterministic FEM solution operator can generalize to previously unseen marginal distributions without retraining, while its accuracy remains dependent on the stochastic parameter ranges covered during identification.

| Test material law | Mean field error (%) | 95th percentile (%) | Tip-mean error (%) | Tip-SD error (%) |
|---|---|---|---|---|
| Gaussian (seen) | 0.653 | 1.640 | 0.387 | 4.427 |
| Lognormal (seen) | 0.559 | 1.341 | 0.153 | 1.054 |
| Gamma (seen) | 0.585 | 1.214 | 0.250 | 1.741 |
| Beta (unseen) | 0.558 | 1.302 | 0.204 | 1.077 |
| Two-point (unseen) | 0.726 | 1.819 | 0.018 | 4.222 |
| Beta (OOD) | 1.149 | 3.618 | 0.959 | 6.095 |
| Two-point (OOD) | 1.780 | 3.815 | 0.430 | 4.084 |

**Table 3**. Physical-coordinate DeepONet accuracy for the 50-element flexural cantilever. OOD denotes simultaneous increases in latent correlation length and coefficient of variation beyond the training ranges.

*7.2.2 Axial bar: explicit Fourier-coordinate transfer*

To investigate cross-law learning directly in Fourier coordinates, we consider a one-dimensional axial bar of length $L = 1$ and constant cross-sectional area $A = 1$, discretized into $N = 16$ equal two-node truss elements. The left end is fixed and a deterministic axial point load $P = 1$ acts at the right end. Young's modulus $E_j$ is constant within element $j$ but varies spatially between elements and across realizations. Under the prescribed tip load, the axial force is the same in every element, and the elongation of element $j$ is $Ph/(AE_j)$, where $h = L/N$. The exact free-nodal displacement is therefore

$$u = G\left[\frac{1}{\mathrm{E}}\right], \quad G_{ij} = \begin{cases} \frac{Ph}{A}, & j \leq i, \\ 0, & j > i. \end{cases} \tag{29}$$

where G is a deterministic lower-triangular compliance-accumulation matrix rather than the inverse of the assembled stiffness matrix. With the complete unitary discrete Fourier transform $F$ of the input modulus and the free-nodal displacement, the corresponding exact Fourier-coordinate map is

$$\hat{u} = F_u \, G\left[\frac{1}{F_E^{-1}\hat{E}}\right] \tag{30}$$

Apparently, this transformation is independent of the probability law assigned to $E$. Although $G$ is linear and deterministic, the mapping from modulus to displacement is nonlinear and therefore, its expansion about a homogeneous reference modulus $E_0$ generates higher-order spectral interactions. Even more, the first-order transfer matrix is generally non-diagonal because the clamped finite domain lacks translation invariance. This reciprocal dependence provides the connection with the higher-order spectral transfer hierarchy described in Section 4.

A full-spectrum Fourier surrogate was trained on the real and imaginary parts of the 16 input Fourier coefficients of the map in the corresponding exact transfer expression and used to predict the corresponding 16 response coefficients. Both identification experiments employed the same multilayer-perceptron architecture, $32 \rightarrow 128 \rightarrow 128 \rightarrow 64 \rightarrow 32$, with 3,000 training and 600 independent validation realizations. In the first experiment, the identification measure comprised Gaussian, lognormal and Gamma correlated modulus fields (1,000 realizations of each). In the second, the same network was trained exclusively on an artificial bounded white-field measure, with independent element moduli $E_j \sim \mathcal{U}(0.5,1.5)$. Both trained models were frozen before evaluation on correlated modulus fields, including previously unseen Beta and two-point marginals and out-of-distribution (OOD) tests at $\ell_c/L = 0.45$ and $\mathrm{CoV} = 0.30$. The Fourier-coordinate neural architecture is shown in Figure 8.

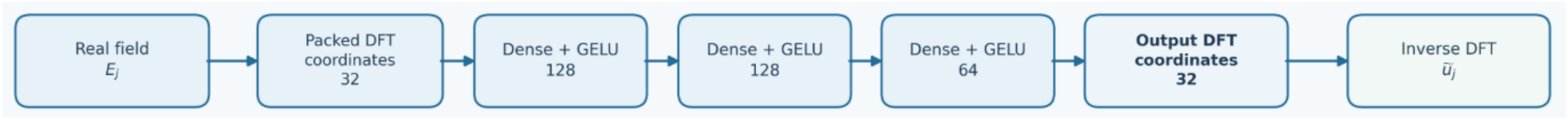

**Figure 8.** Full Fourier-coordinate multilayer perceptrons mapping input modulus coefficients to response coefficients

Table 4 compares the two Fourier-coordinate identification measures. For the Gaussian, Lognormal, and Gamma laws included in training, the three-law surrogate gives lower realization-wise field errors (0.038–0.040%) than the white-only surrogate (0.214–0.358%). The white-only model remains competitive for unseen Beta and two-point fields, with errors of 0.200% and 0.180%, respectively, while the OOD results are nonuniform: the three-law model performs better for Beta fields and the white-only model for two-point fields. The displacement-moment results show the same pattern. The three-law model gives the most accurate mean fields for the Gaussian, Lognormal, Gamma, and unseen Beta ensembles, whereas SD-field errors are consistently larger, particularly for OOD inputs. Thus, broadband artificial excitation is feasible, but its effectiveness depends on the test law and response statistics.

| **Correlated test law** | **Three-law identification: mean field error (%)** | **White-only identification: mean field error (%)** | **White-only identification: mean error (%)** | **White-only identification: SD error (%)** |
|---|---|---|---|---|
| Gaussian (seen by three-law model) | 0.040 | 0.214 | 0.146 ± 0.033 | 2.700 ± 0.618 |
| Lognormal (seen by three-law model) | 0.038 | 0.358 | 0.168 ± 0.056 | 1.541 ± 0.156 |
| Gamma (seen by three-law model) | 0.040 | 0.220 | 0.159 ± 0.035 | 1.776 ± 0.667 |
| Beta (unseen) | 0.205 | 0.200 | 0.175 ± 0.120 | 2.344 ± 0.508 |
| Two-point (unseen) | 0.234 | 0.180 | 0.121 ± 0.051 | 0.968 ± 0.362 |
| Beta (OOD) | 0.583 | 0.832 | 2.898 ± 0.611 | 5.614 ± 0.730 |
| Two-point (OOD) | 0.922 | 0.402 | 3.313 ± 0.567 | 4.963 ± 0.948 |

**Table 4**. Mean relative displacement-field errors for axial-bar Fourier surrogates trained with three correlated modulus laws or an artificial bounded white-field measure.

### *7.3 Statically indeterminate clamped–clamped beam: training measures and stochastic response fidelity*

Both preceding systems are statically determinate under their applied tip loads. Their internal axial force or bending-moment diagram is fixed by equilibrium, regardless of the realized stiffness. Consequently, their exact solution depends linearly on the compliance $1/E$, despite being nonlinear in $E$. A clamped-clamped beam under distributed loading is therefore a more demanding extension: its reaction moments change with the entire spatial stiffness realization, and the response is nonlinear even as a functional of the compliance field.

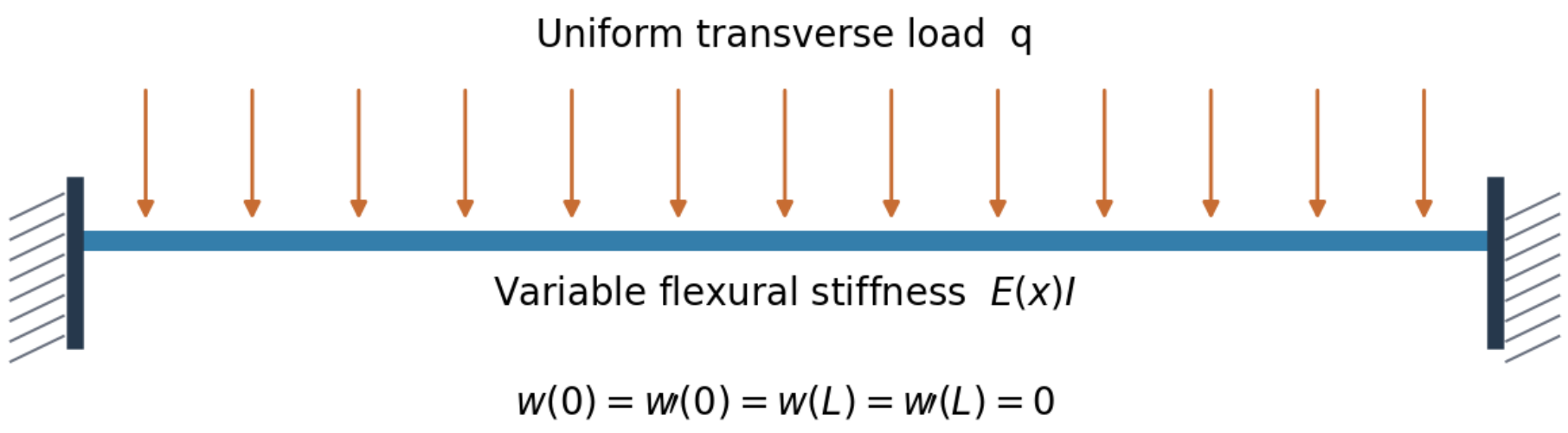

**Figure 9**. Statically indeterminate, clamped-clamped Euler–Bernoulli beam under uniform transverse loading. The reaction moments depend on the realized spatial distribution of flexural compliance.

#### *7.3.1 Deterministic reference solution and stochastic inputs*

Consider the prismatic Euler–Bernoulli beam of Figure 9 with length $L$, constant second moment of area $I$, a spatially variable positive modulus $E(x)$ and a constant transverse load $q$. Both ends are clamped, so $w(0) = w'(0) = w(L) = w'(L) = 0$. In the adopted sign convention, equilibrium and curvature read

$$M_E(x) = A_E + B_E x + 1/2\, qx^2, \qquad w''(x) = \frac{M_E(x)}{IE(x)}. \tag{31}$$

With $C(x) = 1/E(x)$, the exact displacement field is written as

$$w(x;E) = \frac{1}{I}\int_0^x (x-s)\,[A_E + B_E s + 1/2\, qs^2]C(s)\,ds \tag{32}$$

where the two unknown coefficients $A_E$ and $B_E$ are determined from the end conditions as

$$\begin{bmatrix} \int_0^L C(s)\,ds & \int_0^L s\,C(s)\,ds \\ \int_0^L (L-s)\,C(s)\,ds & \int_0^L (L-s)\,sC(s)\,ds \end{bmatrix} \begin{bmatrix} A_E \\ B_E \end{bmatrix} = -\frac{q}{2} \begin{bmatrix} \int_0^L s^2\,C(s)\,ds \\ \int_0^L (L-s)\,s^2 C(s)\,ds \end{bmatrix}. \tag{33}$$

Unlike the statically determinate examples, the coefficients $A_E$ and $B_E$ involve the inverse of a compliance-dependent two-by-two matrix. This creates nonlocal nonlinear coupling across the modulus field, independently of any probabilistic assumptions.

In this study $L = q = I = 1$, while Young's modulus is considered piecewise constant over 32 equal discretized segments. The compliance integrals are evaluated analytically segment by segment therefore, the exact-response reference is free of finite-element solution error, although it refers to this specific 32-segment material discretization. Responses are sampled at 32 spatial locations $x_j = jL/32$ $(j = 0,\dots,31)$, including the midspan $x = L/2$. The analytical reference solution enforces zero displacements at the right clamp and therefore they are omitted from the neural-network output vector.

Four identification source families are used. W comprises independent positive bounded moduli $E_e \sim \mathcal{U}(0.4,1.6)$. S comprises exponentially correlated latent-Gaussian-copula fields with $\ell_c/L \in [0.04,0.55]$ and Cov $\in [0.10,0.38]$ and Gaussian, lognormal or Gamma marginals; Gaussian material values are truncated below 0.25 to impose positivity of the modulus. Beta and two-point correlated marginals are excluded from the S training pool and reserved for transfer tests. M comprises random-phase single-tone fields $E(x) = 1 + a\cos(2\pi kx/L + \varphi)$ with $k = 1,\dots,14$ and $a \in [0.10,0.65]$. T comprises the sum of two independently phased cosine components, with individually sampled amplitudes in the range $[0.06,0.31]$. The source families therefore differ not only in spectral structure but also in their amplitude distributions.

### _7.3.2 Learning the Fourier-coordinate transfer_

The Fourier surrogate maps the complete packed real Fourier coordinates of the 32-point modulus vector to those of the 32-point response. The neural architecture is 32–128–128–32, and all output frequencies are retained. Six excitation mixtures are examined: W only; S only; M only; equal W/S; 33% W + 33% S + 34% M; and equal W/S/M/T. Each independent run uses 1,300 training fields and 240 validation fields, with two independently initialized runs per mixture. The fixed test sets contain 320 examples from each W, S, M and T family. The corresponding full Fourier-coordinate network is shown in Figure 10.

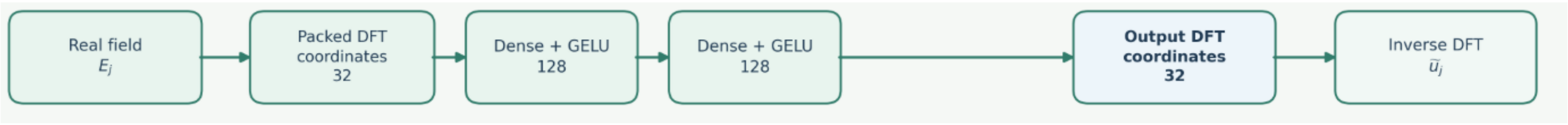

**Figure 10**. Full Fourier-coordinate multilayer perceptrons mapping input modulus coefficients to response coefficients

Table 5 compares the effect of training-source composition on the Fourier-coordinate surrogate. Its Macro field-error column is the unweighted average of the four family-wise errors, not a distribution-free performance guarantee. White-only identification gives the lowest macro-average in this experiment (0.93%), closely followed by the four-way W/S/M/T mixture (0.98%). The four-way mixture improves smoothly correlated and two-tone accuracy relative to white-only identification but gives a larger error on white fields. Monochromatic-only training is accurate on its own excitation family (0.30%) but generalizes poorly to white and smooth-correlated fields (43.64% and 82.93%), illustrating the need to sample a sufficiently rich range of nonlinear mode interactions.

| Training measure | W test (%) | S test (%) | M test (%) | T test (%) | Macro (%) |
|---|---|---|---|---|---|
| White only | 1.22 | 0.98 | 0.93 | 0.57 | 0.93 |
| Smooth only | 4.72 | 0.92 | 2.97 | 1.69 | 2.58 |
| Monochromatic only | 43.64 | 82.93 | 0.30 | 0.86 | 31.93 |
| White / smooth (50/50) | 1.92 | 0.71 | 1.26 | 0.76 | 1.16 |
| White / smooth / mono (33/33/34) | 2.29 | 0.84 | 0.58 | 0.58 | 1.07 |
| Four-way (25% each) | 2.03 | 0.75 | 0.62 | 0.53 | 0.98 |

**Table 5**. Mean relative displacement-field prediction errors (%) of the full Fourier-coordinate surrogate for the clamped–clamped beam, averaged over two training seeds. W: bounded white; S: smooth correlated; M: single-tone; T: two-tone.

To assess the effect of training-set coverage on standard-deviation prediction, the reconstructed clamped–clamped beam operator was trained with balanced white, smooth-correlated, monochromatic, and two-tone ensembles containing 500, 1,500, 4,500, and 9,000 fields. The models were evaluated on the same independent shifted-Beta test ensemble. The mean relative SD-field error was 5.603%, 5.995%, 4.430%, and 0.569%, respectively, with the final value accompanied by a seed-to-seed spread of 0.151%. Thus, increasing both the number and spectral diversity of the training fields reduced the SD error by approximately 90% relative to the 500-field experiment and substantially improved reproducibility across training seeds. The nonmonotonic behavior at the smaller training sizes reflects finite-sample and optimization variability, whereas the 9,000-field model provides a clear improvement in both accuracy and stability. The corresponding convergence trend is shown in Figure 7.

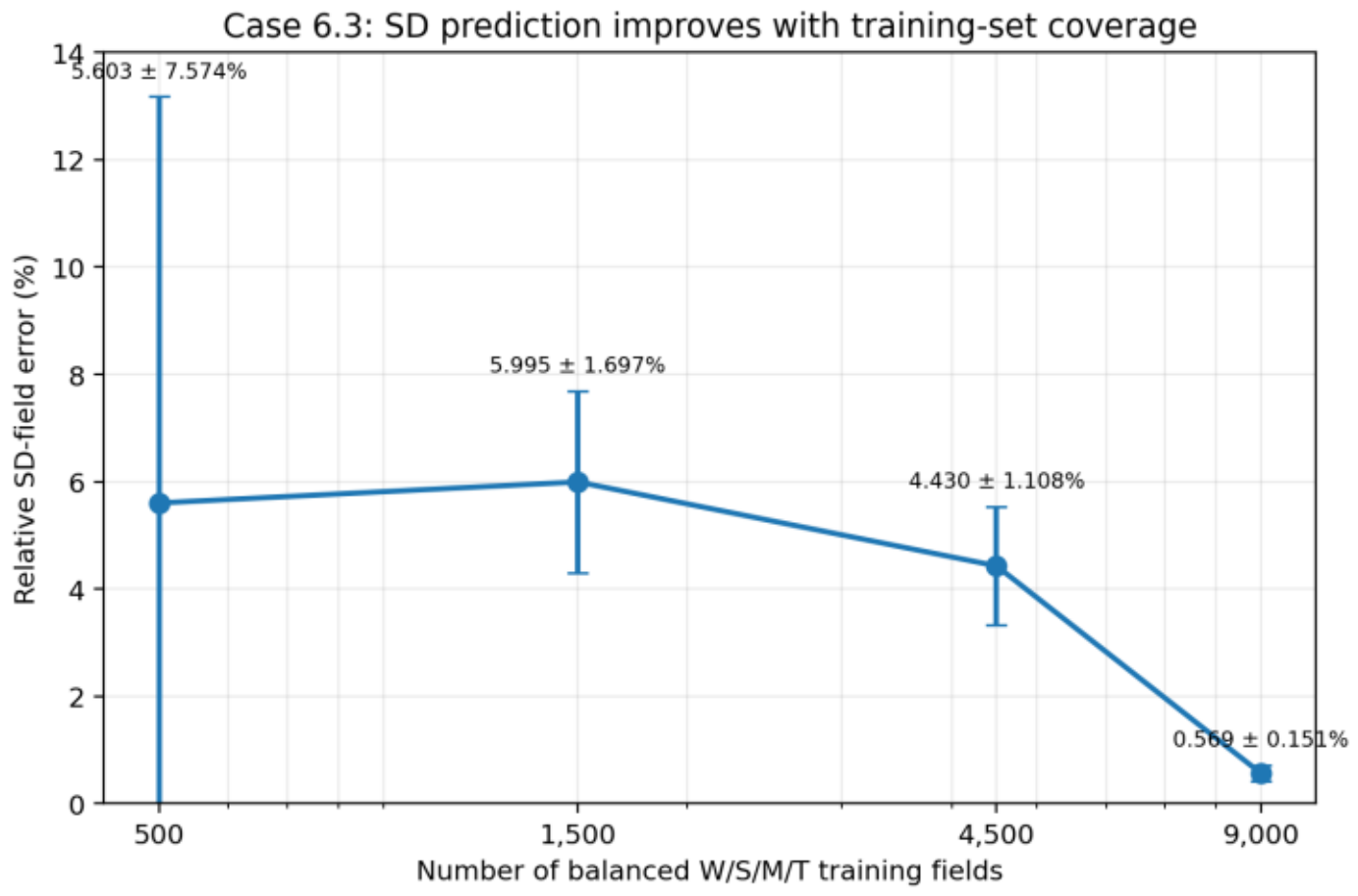

**Figure 7.** Convergence of the relative standard-deviation-field error with training-set size

*7.3.3 Fixed-reference PCE transfer*

A separate coefficient-to-coefficient experiment uses the same clamped–clamped reference mechanics but a fixed two-dimensional uniform germ $\xi = (\xi_1, \xi_2)$, with independent components uniformly distributed in $[-1,1]$. The input is parameterized as

$$E(x, \xi) = 1 + A_1(x)\xi_1 + A_2(x)\xi_2, \tag{34}$$

For each stochastic input model, the amplitude functions $A_1(x)$ and $A_2(x)$ are fixed. Realizations of that model are obtained by varying only the two independent uniform germ variables $\xi_1$ and $\xi_2$. Across training models, however, the spatial amplitude patterns themselves are resampled. Their magnitudes are bound to maintain a strictly positive modulus for all admissible germ values so that $|A_1(x)| + |A_2(x)| < 1$ everywhere. These amplitude families control which spatial structures and stochastic directions are excited. In the white-shaped class (W), amplitude values are generated independently over the 32 segment locations before any positivity rescaling, while in the smooth-shaped class (S), neighboring amplitudes are spatially correlated. A representative monochromatic model (M) uses $A_1(x) = a\cos(2\pi kx/L+\varphi)$ and $A_2(x) = 0$, whereas a representative two-tone model (T) uses two cosine amplitude functions with independently selected frequencies and phases. Unlike M, T can activate both $\xi_1$ and $\xi_2$. These descriptors refer to the spatial amplitude functions, not to changes in the uniform reference-germ law.

The response is represented in a fixed orthonormal total-degree-four Legendre basis in ($\xi_1$, $\xi_2$). This basis contains (4 + 2 choose 2) = 15 polynomials at each of the 32 response locations. The corresponding learned coefficient map has 64 inputs, two spatial amplitude vectors of length 32 each and 480 outputs corresponding to 15 response coefficients at 32 locations. The operator is approximated by the 64–192–192–480 multilayer perceptron shown in Figure 11. Reference response coefficients are obtained by applying 11×11 Gauss–Legendre quadrature to the analytical clamped–clamped beam response. Each experiment uses 450 training stochastic models, 90 validation models and 100 independent held-out models for each W/S/M/T class, with two training seeds for each mixture. Errors are computed from reconstructed physical response fields.

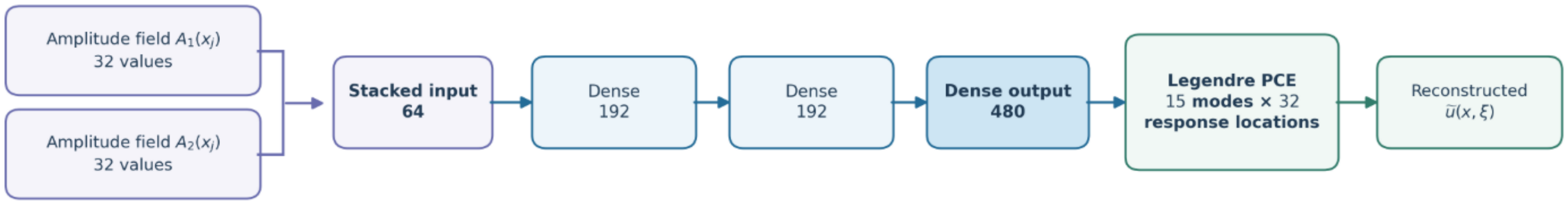

**Figure 11.** Fixed-reference polynomial-chaos surrogate for the clamped–clamped beam, mapping two spatial amplitude functions into response coefficients through a 64–192–192–480 network, with 15 Legendre modes at each of 32 spatial locations.

| PCE training measure | W-shaped (%) | S-shaped (%) | M-shaped (%) | T-shaped (%) | Macro (%) |
|---|---|---|---|---|---|
| White-shaped only | 0.42 | 0.84 | 0.75 | 0.67 | 0.67 |
| Smooth-shaped only | 0.54 | 0.44 | 0.88 | 0.80 | 0.67 |
| Monochromatic only | 7.05 | 11.69 | 0.06 | 9.42 | 7.05 |
| White / smooth / mono (33/33/34) | 0.48 | 0.56 | 0.22 | 0.72 | 0.50 |
| Four-way (25% each) | 0.45 | 0.56 | 0.24 | 0.55 | 0.45 |

**Table 6**. Reconstructed physical-field errors (%) for the degree-four, two-germ Legendre-PCE surrogate trained with different amplitude-shape mixtures (two independent runs per mixture)

Table 6 shows that the equal four-way W/S/M/T training mixture attains the smallest average reconstructed-field error (0.45%), followed by the three-way W/S/M mixture (0.50%). White-shaped-only and smooth-shaped-only training both give 0.67%. Monochromatic-only training reproduces its own one-tone test class accurately (0.06%) but generalizes poorly to white-shaped, smooth-shaped and two-tone test models (7.05%, 11.69% and 9.42%, respectively), leading to a 7.05% macro-average. This result is consistent with missing excitation of the second germ direction and its associated mixed chaos modes, rather than with a change in the underlying deterministic mechanical operator.

*7.3.4 Second order validation*

Figure 12 compares the effect of the training-measure mixture on the baseline Fourier and PCE coefficient learners. Each value is the equal-weight W/S/M/T macro-average formed within a seed. Bars and error bars show the mean and sample standard deviation across two seeds in Log error axes, used to display the wide range of errors. In both representations, the mean response is reproduced more accurately than the standard-deviation field. The difference is particularly pronounced for the PCE operator: its mean coefficient is learned accurately, whereas errors in the nonconstant PCE coefficients accumulate when the standard-deviation field is reconstructed. The Fourier learner shows a more stable response across the tested training mixtures, while the PCE learner is more sensitive to the coverage of the stochastic amplitude functions and the accuracy of its fluctuation coefficients. This comparison demonstrates that small realization-wise or mean-field errors do not guarantee accurate response variability.

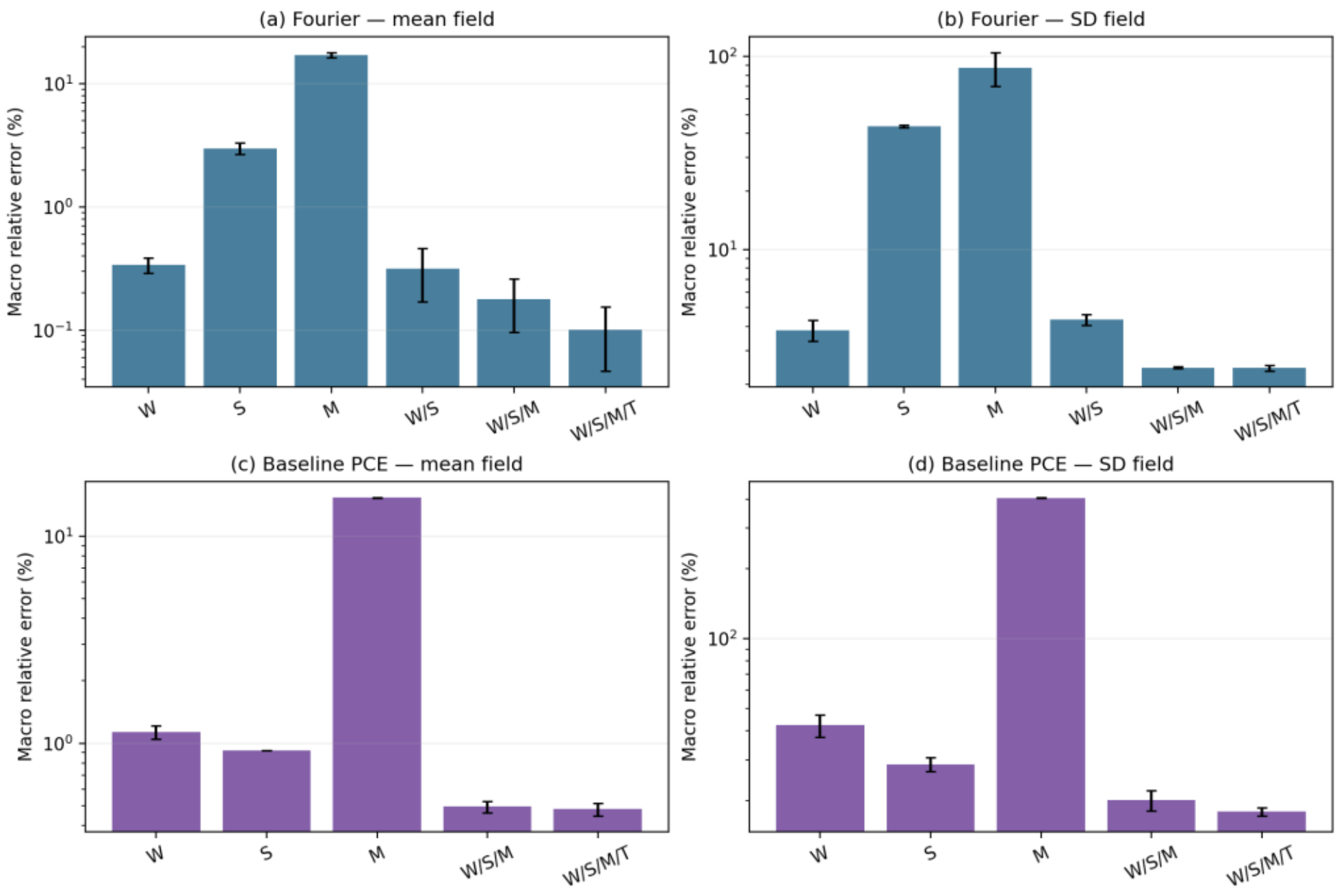

**Figure 12**. Mean-field and SD-field errors for reconstructed clamped-beam training-mixture studies: (a,b) Fourier-coordinate learners and (c,d) baseline PCE coefficient learners.

**Improved PCE operator.** The PCE operator therefore requires targeted improvement through separate normalization of the mean and nonconstant coefficients, an explicit SD-field loss, physics-based coefficient information, symmetry constraints, and broader training coverage, augmentation and inference projection. Training was extended from 180 to 1,200 epochs. The first experiment used the same 450 training amplitude models as the baseline, while a second used 2,400 models to increase coverage of the amplitude-function space. The core architecture of the improved fixed-reference PCE operator is given in Figure 13.

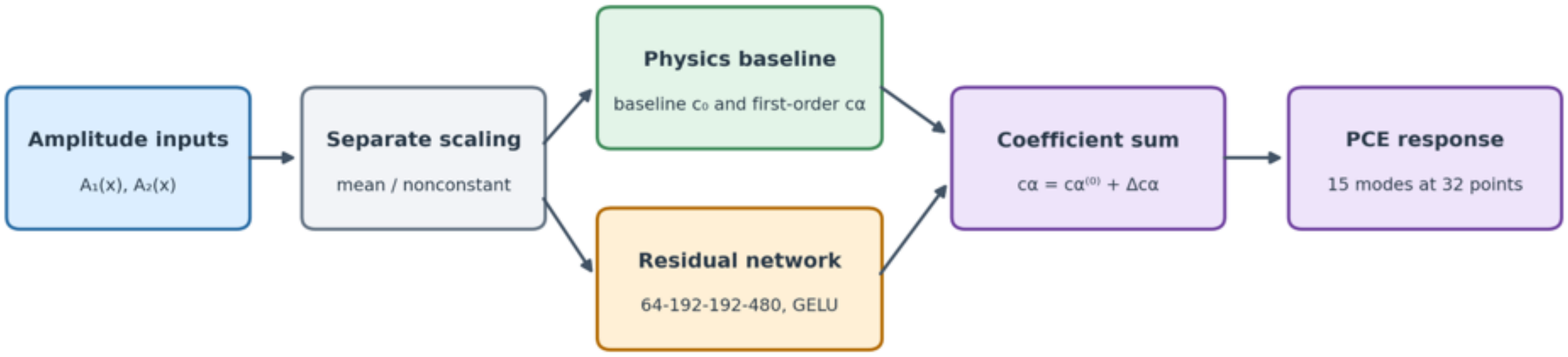

**Figure 13.** Core architecture of the improved fixed-reference PCE operator. The amplitude functions are separately scaled, combined with physics-based baseline coefficients and residual neural corrections, and mapped to the full PCE response.

Figure 14 summarizes the resulting improvement across the white-shaped, smooth-shaped, monochromatic, and two-tone test families. The macro SD field error decreases from 17.940% for the baseline learner to 3.400 ± 0.167% with 450 training models and to 1.091 ± 0.026% with 2,400 models. An independent test gives 3.292 ± 0.176% and 1.097 ± 0.012%, respectively. These results show that the SD error is controlled primarily by the accuracy and training coverage of the learned nonconstant PCE coefficients, rather than by quadrature or by the degree-four representation itself

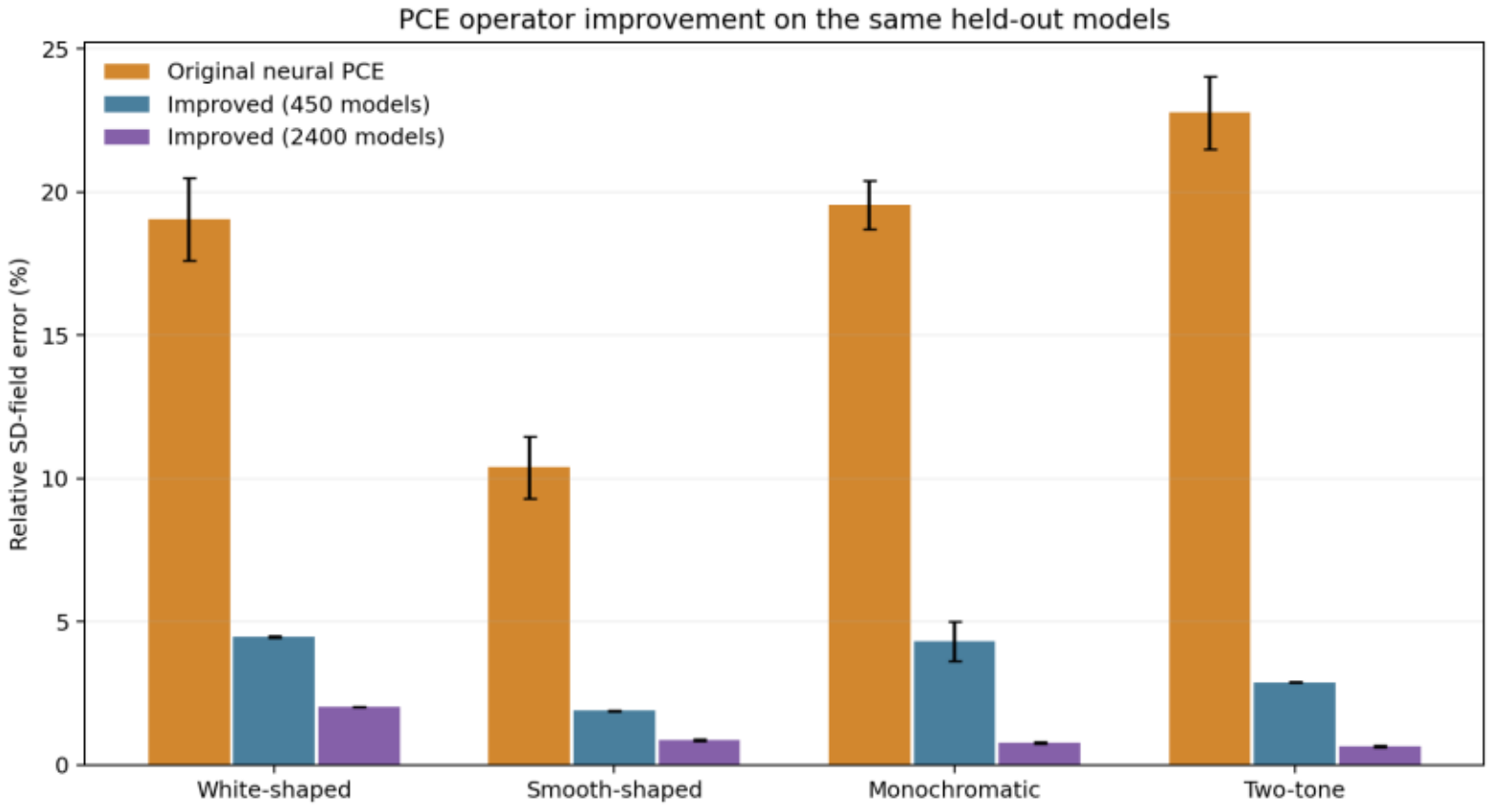

**Figure 14**. Improved reconstructed two-germ PCE operator: SD-field errors on the same held-out W/S/M/T amplitude models

*7.3.5 Midspan probability density validation*

To assess distributional fidelity, probability-density functions (PDFs) of the midspan displacement were compared with the analytical reference using identical input realizations. These results are presented in Figure 15. The Fourier-coordinate surrogates reproduce the reference Beta-law PDF reasonably well for both the balanced W/S/M/T and white-only training measures, with only modest changes in the peak and right-hand tail. This indicates that accurate realization-wise field prediction can transfer to the response distribution when the Fourier coefficient map is adequately identified.

The PCE comparison reveals a stronger dependence on coefficient-learning quality. In the smooth test model, the original neural PCE produces a substantially displaced and overly concentrated PDF, whereas the direct degree-four PCE and the improved neural PCE trained with 2,400 amplitude models closely follow the analytical reference over the main body and tail of the distribution. The improved PCE therefore restores distributional accuracy by reducing errors in the nonconstant coefficients; the improvement is attributable to the coefficient-learning strategy and training coverage, not to changing the PCE degree or increasing the quadrature order.

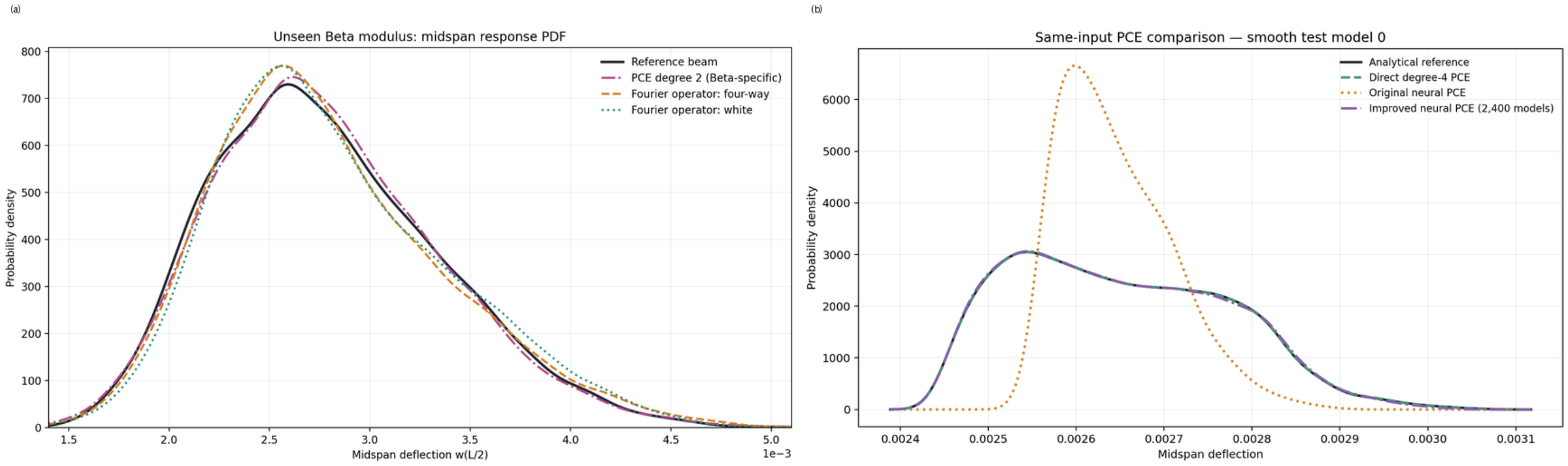

**Figure 15**. Midspan displacement PDFs. (a) Unseen correlated Beta-modulus test comparing the analytical reference with Fourier surrogates trained using the balanced W/S/M/T and white-only measures, together with a degree-two PCE reference. (b) Smooth test model comparing the analytical reference, direct degree-four PCE, original neural PCE, and improved neural PCE trained with 2,400 amplitude models. The improved PCE closely reproduces the reference PDF.

### 7.3.6 Convergence with training-set size and network architecture

A convergence study was performed for the complete Fourier-coordinate surrogate of the same 32-segment clamped–clamped beam, using a fixed balanced W/S/M/T identification measure. Training sets contained 500, 1,500, 4,500, or 9,000 realizations, and all models were evaluated on the same independent 60,000-realization correlated Beta test ensemble. An approximately 1% midspan-deflection threshold was calibrated beforehand from 50,000 exact responses; the independent exact test gave an exceedance probability of 1.095%.

The results of this study are presented in Figure 16. From this figure it can be seen that increasing the training set reduces the mean full-field error from 1.201% to 0.320%, while the mean absolute exceedance-probability error decreases from 0.321 to 0.204 percentage points. The tail estimate therefore improves more slowly and less regularly than the field prediction. At a fixed budget of 4,500 training fields, the 3 × 128 network gives the lowest field error (0.381%) and exceedance error (0.198 percentage points) among the tested architectures, compared with 0.658% and 0.489 percentage points for the 2 × 64 network. These results show that training coverage and architecture both affect performance, but lower field error does not guarantee accurate rare-event probabilities. Broader testing is required to establish reliable tail convergence.

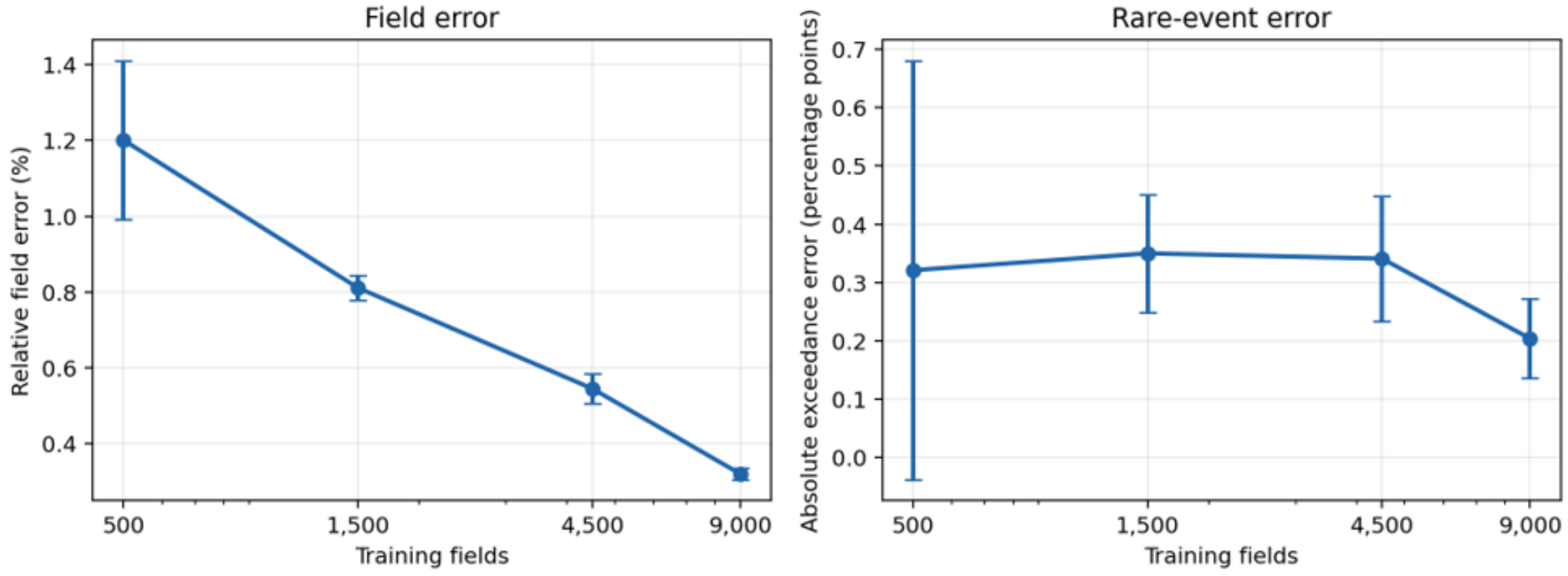

**Figure 16**. Convergence of the Fourier surrogate with training-set size under the fixed balanced W/S/M/T identification measure. Markers and error bars show three-seed means and standard deviations for the relative field error and the absolute error in the approximately 1% midspan-deflection exceedance probability.

### 7.4 Two-dimensional stochastic elasticity: FEM-based supervised cross-law validation

Consider the two-dimensional plane-stress cantilever in Figure 17. The nondimensional domain $\Omega = (0, 2) \times (0, 1)$ has unit thickness and is clamped on its left edge. A uniform downward traction $t_y = -0.002$ acts on the right edge; the remaining boundaries are traction-free. Infinitesimal strains and a constant Poisson ratio $\nu = 0.30$ are assumed, whereas the strictly positive Young's modulus $E(x, y)$ varies spatially.

The displacement reference is computed using 16 × 8 bilinear four-node quadrilateral (Q4) elements with 2×2 Gauss integration. The 128 elementwise moduli constitute the input vector. With 153 mesh nodes and both displacement components fixed on the nine left-edge nodes, the system matrix contains 288 free degrees of freedom. A sparse FEM solution provides the reference displacement for each material realization.

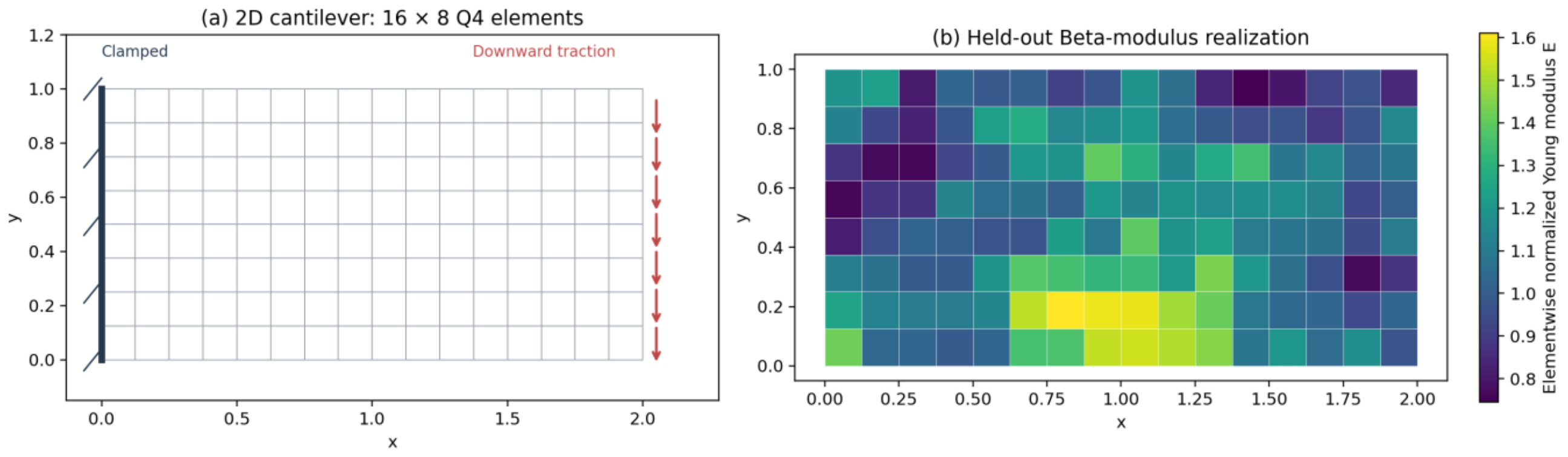

**Figure 17**. Two-dimensional cantilever used in the supervised cross-law study: (a) 16 × 8 Q4 reference mesh, clamped left boundary and right-edge downward traction; (b) an illustrative correlated, held-out Beta-modulus realization. Modulus values are constant within each element.

The correlated training and test fields are generated from standardized latent Gaussian fields with anisotropic covariance $C_{ij} = \exp\{-[(\Delta x/\ell_x)^2 + (\Delta y/\ell_y)^2]^{1/2}\}$. The baseline correlation lengths are $\ell_x = 0.32$ and $\ell_y = 0.22$, with mean modulus normalized to one and coefficient of variation CoV = 0.18. Gaussian, lognormal and Gamma material marginals form the correlated mixed-law identification measure. Beta and two-point marginals are withheld for cross-law testing. Marginals are obtained through transformations of a common underlying Gaussian dependence model, with their actual covariance being modified after nonlinear translation. As in previous cases, modulus positivity is enforced.

For comparison purposes, a second identification measure samples independent bounded element moduli, $E_e \sim U(1-3\mathrm{CoV},\ 1+\sqrt{3}\mathrm{CoV})$. The stronger out-of-distribution (OOD) tests use the previously unseen Beta and two-point marginals while simultaneously changing the correlation lengths to $(\ell_x, \ell_y) = (0.80, 0.50)$ and increasing CoV to 0.30. Thus, the OOD tests combine marginal-law, spatial-correlation and variability shifts rather than isolating one factor.

### 7.4.2 Supervised operator architecture and identification protocol

A fixed-mesh physical-coordinate multilayer perceptron (MLP), approximates the mapping $E \mapsto u$. Its architecture is 128–128–128–288: the 128 modulus inputs pass through two hidden fully connected layers of width 128 with GELU activation and a linear 288-component output layer. The output represents all unconstrained displacement degrees of freedom, so essential boundary conditions are respected by construction through Dirichlet elimination. Let $u_0 = K(1)^{-1}f$ denote the homogeneous-modulus displacement and $U_0 = \max_a |(u_0)_a|$ a fixed displacement scale. The model is trained to predict a correction to the normalized reference $v_0 = u_0/U_0$, rather than to reconstruct the entire displacement from zero:

$$v_\theta(E) = v_0 + \delta v_\theta(E), \quad v_{FEM}(E) = \frac{K(E)^{-1}f}{U_0} \tag{38}$$

The architecture of the approximate operated is depicted in Figure 18. The supervised objective minimizes the squared error in the normalized free displacement components. Reference FEM displacement vectors are generated only for the training/validation realizations; after learning, the frozen model evaluates new modulus fields without a FEM solve.

$$\mathcal{L}_{\text{sup}}(\theta) = \frac{1}{N}\sum_{i=1}^{N} \|v_\theta(E_i) - v_{FEM}(E_i)\|_2^2 \tag{39}$$

Each identification measure contains 720 training and 120 independent validation realizations. Four independent training/data seeds are used for each identification strategy. All runs use AdamW with initial learning rate $10^{-3}$ and at most 320 epochs. The best validation checkpoint is retained. Each frozen model is evaluated on the same 110 independently generated FEM reference fields per test law. The four-seed standard deviations measure run-to-run variability, not sampling confidence intervals.

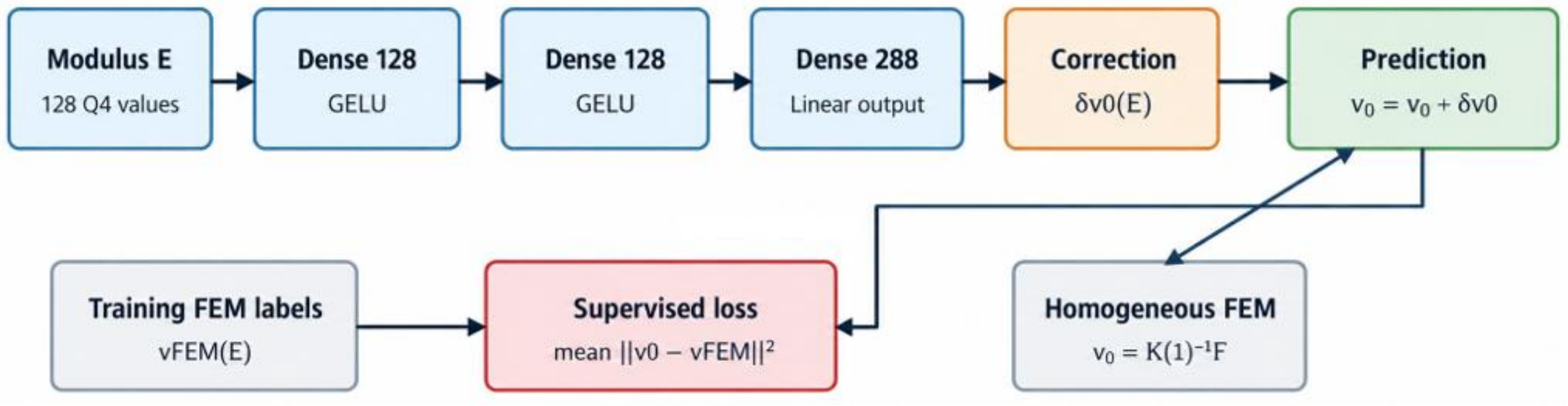

**Figure 18**. Supervised identification architecture for the 2D cantilever.

*7.4.3 Cross-law and out-of-distribution prediction accuracy*

Table 7 displays three distinct regimes. For the seen Gaussian, lognormal and Gamma marginals, mixed-law training achieves very similar mean errors (0.395%, 0.373% and 0.368%), whereas white-only training gives 0.705–0.726%. This is consistent with better approximation on the correlated training ensemble, rather than a universal advantage of mixed-law identification.

| Test material law | Mixed-law training (%) | White-only training (%) |
|---|---|---|
| Gaussian | 0.395 ± 0.015 | 0.705 ± 0.024 |
| Lognormal | 0.373 ± 0.007 | 0.726 ± 0.014 |
| Gamma | 0.368 ± 0.016 | 0.725 ± 0.035 |
| Beta (unseen) | 0.384 ± 0.015 | 0.683 ± 0.045 |
| Two-point (unseen) | 0.476 ± 0.023 | 0.403 ± 0.023 |
| Beta OOD | 0.894 ± 0.043 | 3.129 ± 0.509 |
| Two-point OOD | 1.341 ± 0.073 | 3.350 ± 0.553 |

**Table 7**. Mean relative full-field displacement prediction errors (%), four-seed mean ± standard deviation; each test law comprises 110 *commonly* held-out FEM realizations. Mixed denotes Gaussian/lognormal/Gamma correlated identification; white denotes independent bounded elementwise identification. Beta and two-point marginals are unseen in training.

For the withheld Beta marginal, the mixed-law model reaches 0.384 ± 0.015%, versus 0.683 ± 0.045% for white-only identification. The unseen two-point law reverses the ranking: white-only identification achieves 0.403 ± 0.023%, compared with 0.476 ± 0.023% for mixed-law training. Therefore, the evidence supports transfer without retraining but does not support a claim that a single identification distribution is best for every downstream law. The two-point outcome is also consistent with the different material configurations sampled by an artificial bounded-white measure; the experiment does not identify a unique causal mechanism.

The combined OOD shifts yield a clearer separation: the mixed-law model gives 0.894 ± 0.043% for Beta and 1.341 ± 0.073% for two-point fields, while white-only training gives 3.129 ± 0.509% and 3.350 ± 0.553%, respectively. Relative to their own baseline cross-law errors, both models degrade in these shifted regimes, but the white-only model is markedly more sensitive. The OOD comparison changes multiple input statistics simultaneously, so the contributions of marginal transformation, correlation length and CV cannot be apportioned individually from this experiment.

*7.4.4 Higer moments and edge deflection probability density*

Figure 19 adds a moment-level comparison to the realization-wise errors in Table 7. It reports errors in the predicted mean and pointwise standard-deviation fields for the same paired FEM test samples, comparing mixed-law and white-only identification under baseline conditions and combined out-of-distribution shifts. Across the seen Gaussian, lognormal and Gamma marginals, mixed-law training generally gives lower mean-field and standard-deviation-field errors. For the baseline held-out two-point law, white-only training performs better, consistent with the reversal already noted for realization-wise error in Table 7.

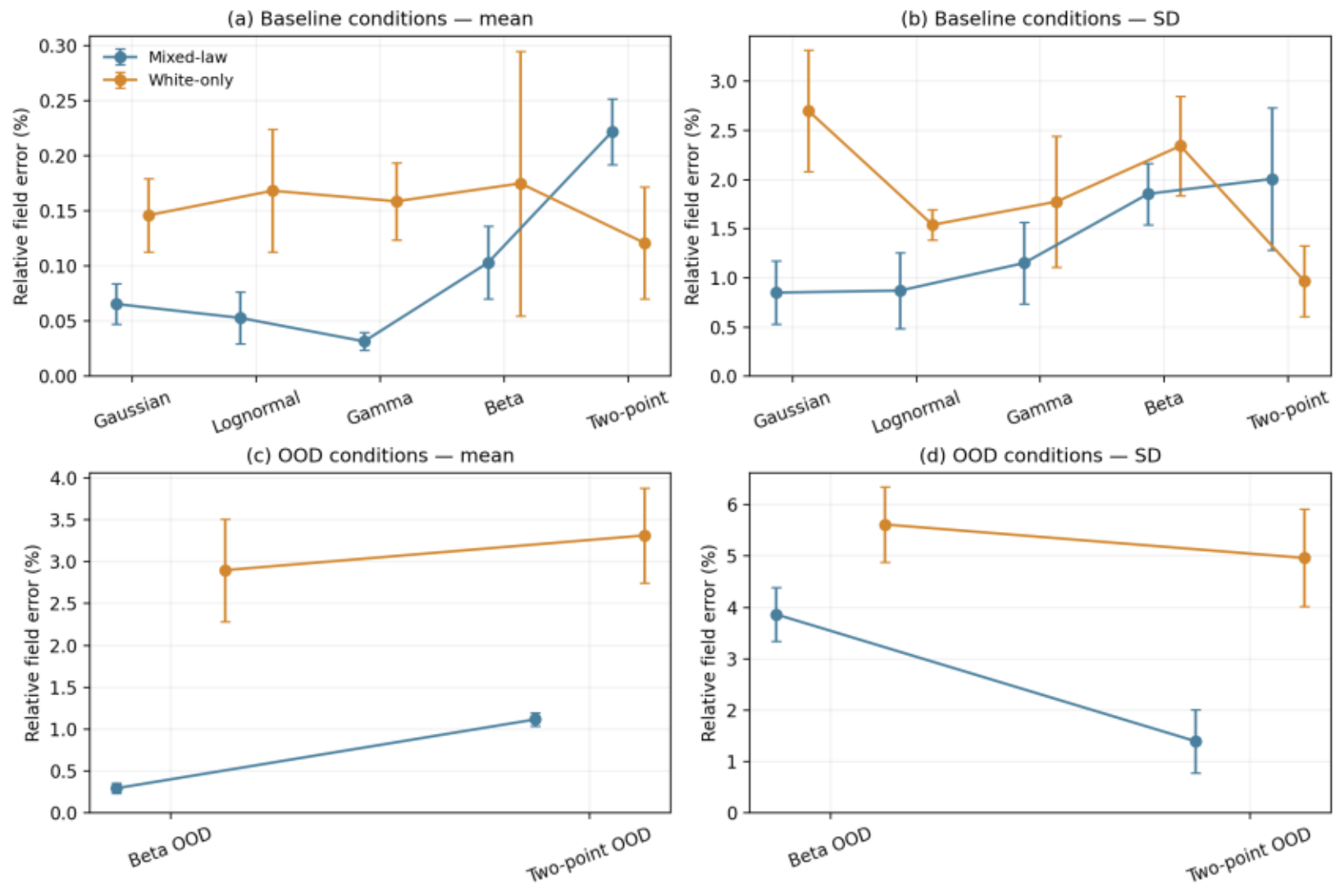

**Figure 19**. Two-dimensional FEM moment errors for mixed-law and white-only supervised identification: (a,b) baseline correlation/variability conditions and (c,d) combined OOD shifts.

The added moment errors make the distinction between field accuracy and response-variability accuracy apparent. For the withheld Beta marginal, mixed-law identification gives a mean-field error of 0.103 ± 0.033% and an SD-field error of 1.856 ± 0.313%. The corresponding white-only errors are 0.175 ± 0.120% and 2.344 ± 0.508%. Thus the displacement standard deviation is reproduced less accurately in relative terms than the mean field, despite sub-percent realization-wise displacement errors.

A separate assessment examines the positive downward displacement at the midpoint of the loaded right edge for an unseen correlated Beta material law. Thresholds at the empirical 95th and 99th percentiles were calibrated using 10,000 independent double-precision FEM responses. The thresholds were then fixed before both frozen supervised models were evaluated on the same 16,000 newly generated modulus fields; corresponding FEM responses establish a paired reference. On the independent 16,000-realization FEM test set, the exact reference exceedance frequencies were 5.281% for the nominal 5% threshold and 1.119% for the nominal 1% threshold. The latter corresponds to 179 exceedances; The corresponding empirical rates and confidence considerations are reported in Table 8.

| Threshold and FEM rate | Mixed supervised (%) | White supervised (%) |
|---|---|---|
| Nominal 5%; FEM 5.281% | 5.745 ± 0.067 | 5.128 ± 0.268 |
| Nominal 1%; FEM 1.119% | 1.295 ± 0.033 | 0.814 ± 0.151 |

**Table 8**. Empirical exceedance rates (%) for the unseen Beta marginal under fixed reference-calibrated displacement thresholds; four-seed mean ± between-seed standard deviation. All four-seed predictions use the same 16,000 independent test fields.

At the nominal 5% threshold the mixed-law network overestimates the FEM frequency by 0.464 percentage points, while the white-only network underestimates it by 0.153 percentage points. At the more demanding nominal 1% threshold, the corresponding predicted frequencies are 1.295% and 0.814%, deviating by +0.177 and −0.305 percentage points from the 1.119% FEM reference. These differences are significant despite sub-percent full-field errors. Therefore again, accurate field predictions do not necessarily ensure accurate low-probability event estimates.

To further examine whether accurate displacement-field predictions translate into accurate stochastic response distributions, empirical probability densities were also constructed. Figure 20 compares four ensembles: Gaussian fields represented during mixed-law training (6,000 samples), previously unseen correlated Beta fields (16,000 samples), previously unseen correlated two-point fields (6,000 samples), and Beta fields with a simultaneous out-of-distribution increase in correlation lengths and coefficient of variation (6,000 samples). Both surrogates closely follow the finite-element density in the central portion for the Gaussian, Beta and two-point laws. Differences become more visible under the combined OOD shift, particularly for the white-only model, whose density exhibits a more pronounced displacement bias and a reduced spread.

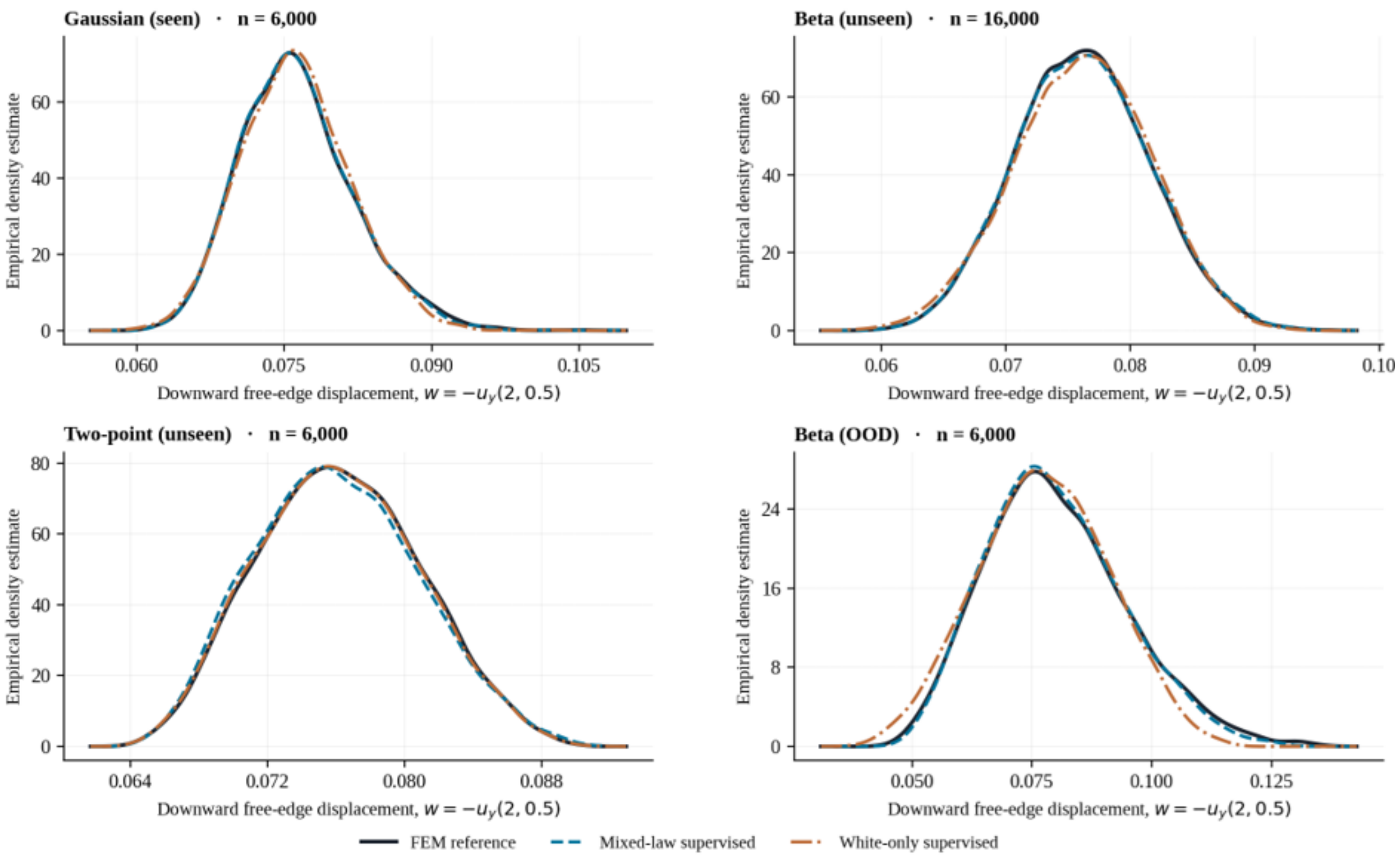

**Figure 19**. Empirical probability density estimates of the downward mid-edge displacement for Gaussian (seen; upper left), Beta (unseen; upper right), two-point (unseen; lower left), and Beta (OOD; lower right) modulus fields.

## *7.5 Discussion of results*

The analytical framework establishes a hierarchy of deterministic operator representations. For linear translation-invariant systems, the Fourier transfer is diagonal and leads to the classical variance relation based on the VRF and input PSD. For linear systems with boundaries, clamps, or spatial nonhomogeneity, the first-order Fourier transfer is generally a full kernel $H_1(q,k)$. Nonlinear mechanics introduce higher-order deterministic transfer kernels $H^n(q;k_1,\ldots,k_n)$, whenever a suitable Volterra-type expansion exists. Neural operators approximate these deterministic input–response mappings and extend the reusable-physics principle of VRF theory to nonlinear and non-diagonal settings.

The numerical studies support this interpretation across several representations and mechanical systems. Fourier coordinates expose spatial spectral interactions, while fixed-reference PCE coordinates describe dependence on a prescribed stochastic germ and provide direct access to response moments. In the realization-to-response tasks, the complete material realization or amplitude description is supplied to the model, so an explicit probability-law label is unnecessary. A law descriptor could become relevant for a different task in which output statistics are predicted from incomplete information about the input field.

The results also show that exact law independence does not imply uniform accuracy of a learned approximation. The quadratic test demonstrates transfer across marginal laws for a prescribed Fourier feature map, while its shifted-spectrum results show sensitivity to changes in spectral content. The cantilever, axial-bar, clamped–clamped beam, and two-dimensional FEM studies further show that cross-law performance depends on the identification measure, training coverage, coordinate representation, and network architecture. Artificial broadband or white excitation can provide effective identification, but no single excitation measure is uniformly optimal across all response laws, correlations, geometries, and error metrics. The two-dimensional FEM benchmark extends this conclusion beyond one-dimensional systems and confirms that the same inverse-stiffness principle applies to coupled plane-stress responses.

The clamped–clamped beam convergence study provides more specific evidence on data size and architecture. With nested balanced training sets, increasing the number of realizations reduced the average displacement-field error. At 4,500 samples, the three-hidden-layer width-128 model outperformed the two-layer width-128 model in both field error and the tested near-1% exceedance estimate. However, increased width was not uniformly beneficial for tail prediction, and the 9,000-sample model still overestimated the exact reference exceedance frequency. These results apply to the specified beam, material law, threshold, and training protocol; they do not establish a general convergence rate. They show that accurate field prediction and accurate rare-event prediction should be evaluated separately, using independent high-accuracy references.

The improved PCE experiment reaches a similar conclusion for response statistics. For the reconstructed two-germ beam benchmark, physics-based initialization, residual learning, symmetry augmentation, and an explicit standard-deviation loss reduced the SD-field error from 17.940% for the baseline coefficient learner to 1.091% for the larger improved model, with 1.097% on an additional independent test. Direct degree-four PCE produced a negligible SD-field error indicating that the dominant baseline limitation was neural coefficient approximation rather than the PCE basis or quadrature. This result supports moment-sensitive objectives and structure-matched architectures for this benchmark; it does not establish general superiority of PCE over Fourier coordinates or guarantee accurate tail probabilities. Overall, the tests indicate that neural-operator architecture, excitation measure, training coverage, and evaluation metric must be selected jointly for the physical and statistical objectives of the problem.

## 7. Conclusions

This work develops and tests a law-invariant operator-learning framework for stochastic mechanics. The central distinction is between the deterministic physical solution operator, which is independent of the probability law assigned to the input, and its learned approximation, whose accuracy depends on the representation, identification measure, training coverage, and model capacity. Complete Fourier and fixed-reference PCE coordinates provide complementary descriptions of the same deterministic input–response map. The analytical, Fourier, finite-element, beam, and two-dimensional benchmarks show that learned operators can transfer across several unseen marginal laws without retraining. The results also show that cross-law transfer is not uniformly accurate: broader or more diverse excitation generally improves coverage, but no single identification measure or architecture is optimal for every system and evaluation metric. Field, moment, and rare-event errors must therefore be assessed separately. The improved PCE experiment further demonstrates that physics-based initialization and explicit treatment of nonconstant coefficients can substantially improve response-variability prediction within the tested stochastic families.

These conclusions are supported by the specified structural models, discretizations, input parameterizations, and training configurations. They do not establish uniform error bounds, general tail-probability convergence, or mesh-independent neural generalization. Future work should develop systematic identification-measure design, scalable multidimensional implementations, controlled comparisons with established uncertainty-quantification methods, and reliability-aware training strategies for accurate low-probability event prediction.

## Declaration of generative AI and AI-assisted technologies

During the preparation of this manuscript, AI-assisted tools were used to support language editing, improve organization and consistency, check cross-references and notation, and prepare schematic illustrations. The tools were not used to generate or modify the underlying data, numerical simulations, equations, or reported scientific results. All technical content, interpretations, citations, figures, and final revisions were reviewed and approved by the authors, who remain fully responsible for the content of the manuscript.